\documentclass[12pt]{article}
\pdfoutput=1
\usepackage[top=1in, bottom=1in, left=1in, right=1in]{geometry}
\usepackage{latexsym}
\usepackage{amssymb, amsfonts, amsmath}
\usepackage{indentfirst}
\usepackage{bbm}
\usepackage{amssymb}
\usepackage{verbatim}
\usepackage{amsmath, amsthm, amssymb}
\usepackage{mathrsfs}
\usepackage{amsfonts}
\usepackage{dsfont}
\usepackage{csquotes}
\usepackage{hyperref}
\usepackage{url}
\usepackage{xcolor}
\usepackage{xpatch}
\usepackage[english]{babel}

\usepackage{graphicx} 
\usepackage[export]{adjustbox}
\usepackage{float}
\graphicspath{ {images/} }

\usepackage[style=phys, articletitle=true, biblabel=brackets, backend=bibtex,
chaptertitle=false, pageranges=false, eprint=true,%
natbib=true, maxbibnames=10, giveninits=true, sorting=none]{biblatex} 

\DeclareFieldFormat[article]{journaltitle}{\mkbibitalic{#1\isdot}} 
\makeatletter
\newrobustcmd{\mkbibfixedbrackets}[1]{%
	\begingroup
	\blx@blxinit
	\blx@setsfcodes
	\bibleftbracket#1\bibrightbracket
	\endgroup}

\renewbibmacro*{doi+eprint+url}{%
	\iftoggle{bbx:doi}
	{\printfield{doi}}
	{}%
	\ifboolexpr{test {\iffieldequalstr{eprinttype}{arxiv}} or test {\iffieldequalstr{eprinttype}{arXiv}}}
	{\setunit{\addspace}\newblock}
	{\newunit\newblock}%
	\iftoggle{bbx:eprint}
	{\usebibmacro{eprint}}
	{}%
	\newunit\newblock
	\iftoggle{bbx:url}
	{\usebibmacro{url+urldate}}
	{}}

\xpatchbibdriver{online}
{\newunit\newblock
	\iftoggle{bbx:eprint}
	{\usebibmacro{eprint}}
	{}}
{\ifboolexpr{test {\iffieldequalstr{eprinttype}{arxiv}} or test {\iffieldequalstr{eprinttype}{arXiv}}}
	{\setunit{\addspace}\newblock}
	{\newunit\newblock}%
	\iftoggle{bbx:eprint}
	{\usebibmacro{eprint}}
	{}}
{}{}

\DeclareFieldFormat{eprint:arxiv}{%
	\mkbibbrackets{%
		\ifhyperref
		{\href{http://arxiv.org/\abx@arxivpath/#1}{%
				arXiv\addcolon
				\nolinkurl{#1}%
				\iffieldundef{eprintclass}
				{}
				{\addspace\UrlFont{\mkbibfixedbrackets{\thefield{eprintclass}}}}}}
		{arXiv\addcolon
			\nolinkurl{#1}%
			\iffieldundef{eprintclass}
			{}
			{\addspace\UrlFont{\mkbibfixedbrackets{\thefield{eprintclass}}}}}}}
\makeatother

\parskip=\medskipamount

\DeclareMathAlphabet{\mathbbmsl}{U}{bbm}{m}{sl}

\newcommand{\cA}{{\cal A}}
\newcommand{\cB}{{\cal B}}

\newcommand{\cD}{{\cal D}}

\newcommand{\cH}{{\cal H}}
\newcommand{\cI}{{\cal I}}

\newcommand{\cN}{{\cal N}}

\newcommand{\cQ}{{\cal Q}}

\newcommand{\cZ}{{\cal Z}}

\def\a{\alpha}

\def\b{\beta}

\def\d{\delta}
\def\e{\epsilon}
\def\f{\phi}
\def\g{\gamma}

\def\l{\lambda}

\def\q{\theta}
\def\r{\rho}
\def\s{\sigma}
\def\t{\tau}

\def\x{\xi}

\def\D{\Delta}
\def\F{\Phi}
\def\J{\Psi}

\def\P{\Pi}
\def\Q{\Theta}

\newcommand{\ve}{\varepsilon}                            

\newcommand{\pa}{\partial}                           

\newcommand{\be}{\begin{equation}}
\newcommand{\ee}{\end{equation}}
\newcommand{\bea}{\begin{eqnarray}}
\newcommand{\eea}{\end{eqnarray}}

\newcommand{\ba}{\begin{array}}
\newcommand{\ea}{\end{array}}

\newcommand\scalemath[2]{\scalebox{#1}{\mbox{\ensuremath{\displaystyle #2}}}}

\def\double #1{#1{\hbox{\kern-2pt $#1$}}}

\newcommand{\bsubeq}{\begin{subequations}}
\newcommand{\esubeq}{\end{subequations}}

\numberwithin{equation}{section}

\begin{document}

\begin{titlepage}
\begin{flushright}
May, 2023
\end{flushright}
\vspace{2mm}

\begin{center}
\Large \bf Grassmann-odd three-point functions of conserved supercurrents in 3D $\cN=1$ SCFT
\end{center}

\begin{center}
{\bf
Evgeny I. Buchbinder and Benjamin J. Stone}

{\footnotesize{
{\it Department of Physics M013, The University of Western Australia\\
35 Stirling Highway, Crawley W.A. 6009, Australia}} ~\\
}
\end{center}
\begin{center}
\texttt{Email: evgeny.buchbinder@uwa.edu.au, \\ benjamin.stone@research.uwa.edu.au}
\end{center}

\vspace{4mm}
\begin{abstract}
\baselineskip=14pt
\noindent 

We consider the analytic construction of three-point functions of conserved higher-spin supercurrents in three-dimensional $\cN=1$ superconformal field theory which are Grassmann-odd in superspace. 
In particular, these include the three-point functions of the supercurrent and flavour currents, which contain the three-point functions of the energy-momentum tensor 
and conserved vector currents at the component level. We present an analytic proof 
for arbitrary superspins that these correlators do not possess a parity-violating contribution. We also prove that the parity-even contribution is unique, 
and exists (under an assumption that is well supported by the computational approach of \href{https://arxiv.org/abs/2302.00593}{arXiv:2302.00593}) for arbitrary superspins. 
The construction of the parity-even sector is shown to reduce to solving a  system of linear homogeneous equations with a tri-diagonal matrix of co-rank one, which we solve explicitly for arbitrary superspins.



\end{abstract}
\end{titlepage}

\newpage
\renewcommand{\thefootnote}{\arabic{footnote}}
\setcounter{footnote}{0}

\tableofcontents
\vspace{3mm}
\bigskip\hrule


\section{Introduction}\label{section1}

In conformal field theory (CFT), the general structure of correlation functions is highly constrained by conformal symmetry.
In particular, the three-point functions of conserved currents such as the energy-momentum tensor, flavour currents and more generally higher-spin currents, 
are fixed up to finitely many independent structures~\cite{Polyakov:1970xd, Schreier:1971um, Migdal:1971xh, Migdal:1971fof,Ferrara:1972cq,Ferrara:1973yt, 
Koller:1974ut, Mack:1976pa, Fradkin:1978pp, Stanev:1988ft,Osborn:1993cr, Erdmenger:1996yc}.
For conformal field theories in three dimensions (3D), it has been proven that the three-point functions of conserved higher-spin currents are constrained up to only three 
independent structures 
\cite{Giombi:2011rz, Maldacena:2011jn, Zhiboedov:2012bm,Giombi:2016zwa,Buchbinder:2022mys}. 
Two of the structures are parity-even (corresponding to free theories), and one is parity-odd (or parity violating), which has been shown to correspond to theories of a 
Chern-Simons gauge field interacting with parity-violating matter 
(see e.g. \cite{Giombi:2011kc, Aharony:2011jz, Closset:2012vp, GurAri:2012is, Aharony:2012nh,Jain:2012qi, Chowdhury:2017vel, 
Skvortsov:2018uru, Inbasekar:2019wdw, Jain:2021gwa, Prakash:2022gvb}). In superconformal field theory (SCFT), three-point functions are further 
constrained.\footnote{The study of correlation functions in superconformal theories has been carried out in diverse dimensions using the group-theoretic approach 
developed in the following publications \cite{Osborn:1998qu, Park:1998nra, Park:1999pd, Park:1999cw, Kuzenko:1999pi, Nizami:2013tpa, Buchbinder:2015qsa, 
Buchbinder:2015wia, Kuzenko:2016cmf, Buchbinder:2021gwu, Buchbinder:2021izb, Buchbinder:2021kjk, Buchbinder:2021qlb, Jain:2022izp, Buchbinder:2021qlb, 
Buchbinder:2022cqp, Buchbinder:2022kmj}.}
For example, in 3D $\cN=1$ superconformal field theory it was shown in \cite{Buchbinder:2021gwu} that there is an apparent tension between supersymmetry and the 
existence of parity-violating structures in three-point functions. In contrast with the non-supersymmetric case, it was shown that parity-odd structures are not found in the 
three-point functions of the energy-momentum tensor and conserved 
vector currents, which were studied using a manifestly 
supersymmetric approach in \cite{Buchbinder:2015qsa,Buchbinder:2015wia,Kuzenko:2016cmf,Buchbinder:2021gwu,Buchbinder:2023fqv}.

The general structure of three-point functions of higher-spin supercurrents in 3D $\cN=1$ SCFT was elucidated 
in \cite{Buchbinder:2023fqv}. Conformal higher-spin supercurrents of superspin-$s$ (integer or half-integer) are defined as totally symmetric spin-tensor superfields,
$\mathbf{J}_{s} \equiv \mathbf{J}_{\a_{1} ... \a_{2s}}(z) = \mathbf{J}_{(\a_{1} ... \a_{2s})}(z)$, and satisfy the conservation equation
\begin{equation} \label{Conserved supercurrent}
	D^{\a_{1}} \mathbf{J}_{\a_{1} \a_{2} ... \a_{2s}}(z) = 0\, ,
\end{equation}
where $D^{\a}$ is the spinor covariant derivative in 3D $\cN=1$ Minkowski superspace. The most important examples of conserved supercurrents in 
superconformal field theory are the flavour current and supercurrent multiplets, corresponding to the cases $s=\tfrac{1}{2}$ and $s = \tfrac{3}{2}$ respectively 
(for a review of the properties of flavour current and supercurrent multiplets in three dimensions, see \cite{Buchbinder:2015qsa,Korovin:2016tsq} and the references there-in). The flavour current multiplet contains a conserved vector current, while the supercurrent multiplet contains the energy-momentum tensor and the supersymmetry current. 
For higher-spin supercurrents, it was shown by explicit calculations up to a high computational bound $(s_{i} \leq 20)$ that the general structure of the three-point 
function $\langle \mathbf{J}^{}_{s_{1}}(z_{1}) \, \mathbf{J}'_{s_{2}}(z_{2}) \, \mathbf{J}''_{s_{3}}(z_{3}) \rangle$ is fixed up to the following form~\cite{Buchbinder:2023fqv}:
\begin{equation}
	\langle \mathbf{J}^{}_{s_{1}} \mathbf{J}'_{s_{2}} \mathbf{J}''_{s_{3}} \rangle = a \, \langle \mathbf{J}^{}_{s_{1}} \mathbf{J}'_{s_{2}} \mathbf{J}''_{s_{3}} \rangle_{E} + b \, \langle \mathbf{J}^{}_{s_{1}} \mathbf{J}'_{s_{2}} \mathbf{J}''_{s_{3}} \rangle_{O} \, ,
	\label{zh1}	
\end{equation}
where $\langle \mathbf{J}^{}_{s_{1}} \mathbf{J}'_{s_{2}} \mathbf{J}''_{s_{3}} \rangle_{E}$ is a parity-even solution, and $\langle \mathbf{J}^{}_{s_{1}} \mathbf{J}'_{s_{2}} \mathbf{J}''_{s_{3}} \rangle_{O}$ 
is a parity-odd solution. 
For the three-point functions which are Grassmann-even (bosonic) in superspace, the existence of the parity-odd solution is subject to the following superspin triangle inequalities:
\begin{align}
	s_{1} \leq s_{2} + s_{3} \, , && s_{2} \leq s_{1} + s_{3} \, , && s_{3} \leq s_{1} + s_{2} \, .
	\label{zh2}	
\end{align}
When the triangle inequalities are simultaneously satisfied there is one even solution and one odd solution, however, if any of the above inequalities
are not satisfied then the odd solution is incompatible with the superfield conservation equations. On the other hand, for the Grassmann-odd (fermionic) 
three-point functions it was shown that the parity-odd solution appears to vanish in general. Despite being limited by computational power to consider superspins $s_{i} \leq 20$, the pattern was clear and we proposed in~\cite{Buchbinder:2023fqv} that these results hold in general. 

The aim of this paper is to study the Grassmann-odd three-point functions analytically for arbitrary superspins. We use a different approach 
to \cite{Buchbinder:2023fqv}, based on a method of irreducible decomposition of tensors. Quite remarkably, we find 
that these three-point functions can be constructed explicitly for arbitrary superspins. For the parity-violating sector we give a completely analytic proof that 
it vanishes for arbitrary superspins. For the parity-even sector we found that its construction is reduced to solving a homogeneous system of linear equations 
with a tri-diagonal matrix of co-rank one, which proves that the parity-even sector is fixed up to a single structure in general. We also found the solution 
to this system for arbitrary superspins, thus obtaining the explicit form of the parity-even contribution.
Our analysis uses one simplifying assumption which is, however, well supported by our computational approach~\cite{Buchbinder:2023fqv}. 
It was noticed in~\cite{Buchbinder:2023fqv} that if the third superspin satisfies the triangle inequality $s_{3} \leq s_{1} + s_{2}$ it is not necessary to impose the supercurrent 
conservation condition at the third point because it is automatically satisfied and does not give any further restrictions. In this paper we assume that this property
continues to hold for arbitrary superspins. However, we should stress that our proof that the parity-odd sector vanishes does not rely on this assumption. 
It is also inconsequential for our analysis of the parity-even sector, as after imposing the conservation conditions for the first two supercurrents we prove that it is already fixed up to an overall coefficient. Since on general grounds we should expect at least one parity-even solution,\footnote{In some cases the parity-even 
solution and, hence, the entire three-point function, vanishes. However, this occurs only when the three-point function is required to be invariant under permutations of superspace points. 
These cases were analysed 
systematically in~\cite{Buchbinder:2023fqv}.} it follows that the conservation condition for the third supercurrent is, indeed, unnecessary. 

The results of this paper are organised as follows. In section \ref{section2} we provide a brief review of the general structure of the three-point functions of conserved currents 
in 3D $\cN=1$ SCFT. In section \ref{section3} we study Grassmann-odd three-point functions which consist of three conserved supercurrents 
of arbitrary half-integer superspins. We show that the construction of both the parity-even and parity-odd sector is governed by a homogeneous system of linear equations with tri-diagonal matrix. By computing the determinants of the tridiagonal matrices for the parity-even and parity-odd sectors, in the former case we prove that the matrix has co-rank one, and hence the parity-even solution is unique for arbitrary superspins. In the latter case, we prove that 
the matrix is non-degenerate meaning that the parity-odd solution vanishes in general. In section \ref{section4} we perform a similar analysis for Grassmann-odd three-point functions consisting of
one fermionic and two bosonic supercurrents. Appendix \ref{AppA} is dedicated to our 3D conventions and notation.

\vfill


\section{Superconformal building blocks and correlation functions}\label{section2}

In this section we will review the essentials of the group-theoretic formalism used to compute three-point correlation functions of primary superfields in 3D $\cN=1$ 
superconformal field theories. For a more detailed review of the formalism and our conventions, 
the reader may consult \cite{Park:1999cw, Buchbinder:2015qsa, Buchbinder:2022mys, Buchbinder:2023fqv}.

Given two superspace points $z_{1}$ and $z_{2}$, we define the two-point functions
\begin{equation}
	\boldsymbol{x}_{12}^{\alpha \beta} = (x_{1} - x_{2})^{\alpha \beta} + 2 \text{i} \theta^{(\alpha}_{1} \theta^{\beta)}_{2} - \text{i} \theta^{\a}_{12} \theta^{\b}_{12} \, ,  
	\hspace{10mm} \theta^{\alpha}_{12} = \theta_{1}^{\alpha} - \theta_{2}^{\alpha} \,. 
	\label{Two-point building blocks 1}
\end{equation}
%
The two-point function $\boldsymbol{x}_{12}^{\alpha \beta} $
can be split into symmetric and antisymmetric parts as follows:
\begin{equation}
	\boldsymbol{x}_{12}^{\a \b} = x_{12}^{\a \b} + \frac{\text{i}}{2} \ve^{\alpha \beta} \theta^{2}_{12} \, , 
	\hspace{5mm} \q_{12}^{2} = \q_{12}^{\a} \q^{}_{12 \, \a} \, , \hspace{5mm} \boldsymbol{x}_{12}^{2} = - \frac{1}{2} \boldsymbol{x}^{\a \b}_{12} \boldsymbol{x}_{12 \a \b} \, ,
	\label{Two-point building blocks 1 - properties 1}
\end{equation}
where the symmetric component
\begin{equation}
	x_{12}^{\a \b} = (x_{1} - x_{2})^{\alpha \beta} + 2 \text{i} \theta^{(\alpha}_{1} \theta^{\beta)}_{2} \, , \label{Two-point building blocks 1 - properties 2}
\end{equation}
is the standard bosonic two-point superspace interval. It is useful to introduce the normalised two-point functions, denoted by $\hat{\boldsymbol{x}}_{12}$,
\begin{align} \label{Two-point building blocks 3}
	\hat{\boldsymbol{x}}_{12 \, \a \b} = \frac{\boldsymbol{x}_{12 \, \a \b}}{( \boldsymbol{x}_{12}^{2})^{1/2}} \, ,
	\hspace{10mm} \hat{\boldsymbol{x}}_{12}^{\a \s} \hat{\boldsymbol{x}}^{}_{21 \, \s \b} = \d_{\b}^{\a} \, . 
\end{align}
From here we can now construct an operator analogous to the conformal inversion tensor acting on the space of symmetric traceless spin-tensors of arbitrary rank. 
Given a normalised two-point function $\hat{\boldsymbol{x}}$, we define the operator
\begin{equation} \label{Higher-spin inversion operators a}
	\cI_{\a(k) \b(k)}(\boldsymbol{x}) = \hat{\boldsymbol{x}}_{(\a_{1} (\b_{1}} \dots \hat{\boldsymbol{x}}_{ \a_{k}) \b_{k})}  \,.
\end{equation}
This object is essential to the construction of correlation functions of primary operators of arbitrary superspins~\cite{Buchbinder:2023fqv}. 

Now given three superspace points, $z_{1}, z_{2}, z_{3}$, we define the three-point building blocks, $\cZ_{k} = ( \boldsymbol{X}_{ij} , \Q_{ij} )$ as follows:
\begin{align} \label{Three-point building blocks 1}
	\boldsymbol{X}_{ij \, \a \b} &= -(\boldsymbol{x}_{ik}^{-1})_{\a \g}  \boldsymbol{x}_{ij}^{\g \d} (\boldsymbol{x}_{kj}^{-1})_{\d \b} \, , 
	\hspace{5mm} \Q_{ij \, \a} = (\boldsymbol{x}_{ik}^{-1})_{\a \b} \q_{ki}^{\b} - (\boldsymbol{x}_{jk}^{-1})_{\a \b} \q_{kj}^{\b} \, ,
\end{align}
where the labels $(i,j,k)$ are a cyclic permutation of $(1,2,3)$. 
They satisfy many properties similar to those of the two-point building blocks (for simplicity we consider $(\boldsymbol{X}_{12},\Q_{12})$)
\begin{subequations}
	\begin{align} 
		\boldsymbol{X}_{12}^{\a \s} \boldsymbol{X}^{}_{21 \, \s \b} = \boldsymbol{X}_{12}^{2} \d_{\b}^{\a} \, , \hspace{5mm} \boldsymbol{X}_{12}^{2} = - \frac{1}{2} \boldsymbol{X}_{12}^{\a \b}  \boldsymbol{X}^{}_{12 \, \a \b} \, ,
	\end{align}
	\begin{equation}
		\boldsymbol{X}_{12}^{2} = \frac{\boldsymbol{x}_{12}^{2}}{\boldsymbol{x}_{13}^{2} \boldsymbol{x}_{23}^{2}} \, , \hspace{10mm}  \Q_{12}^{2} = \Q^{\a}_{12} \Q^{}_{12 \, \a} \, , \label{Three-point building blocks 2}
	\end{equation}
\end{subequations}
and may be decomposed into symmetric and anti-symmetric parts similar to \eqref{Two-point building blocks 1 - properties 1} as follows:
\begin{equation}
	\boldsymbol{X}_{12  \, \a \b} = X_{12  \, \a \b} - \frac{\text{i}}{2} \ve_{\a \b} \Q_{12}^{2} \, , \hspace{10mm} X_{12  \, \a \b} = X_{12  \, \b \a} \, . 
	\label{Three-point building blocks 1a - properties 3}
\end{equation}
The symmetric spin-tensor, $X_{12  \, \a \b}$, can be equivalently represented by the three-vector $X_{12 \, m} = - \frac{1}{2} (\g_{m})^{\a \b} X_{12  \, \a \b}$. 
One may also identify the superconformal invariant
%
\begin{equation}
	\boldsymbol{J} = \frac{\Q_{12}^{2}}{\sqrt{\boldsymbol{X}_{12}^{2}}} =  \frac{\Q_{23}^{2}}{\sqrt{\boldsymbol{X}_{23}^{2}}}= \frac{\Q_{31}^{2}}{\sqrt{\boldsymbol{X}_{31}^{2}}}\,.
\end{equation}
Analogous to the two-point functions, it is also useful to introduce the following normalised three-point building blocks, denoted by $\hat{\boldsymbol{X}}_{12}$, $\hat{\Q}_{12}$:
\begin{align} \label{Normalised three-point building blocks}
	\hat{\boldsymbol{X}}_{12 \, \a \b} = \frac{\boldsymbol{X}_{12 \, \a \b}}{( \boldsymbol{X}_{12}^{2})^{1/2}} \, , \hspace{10mm} \hat{\Q}_{12}^{\a} = \frac{ \Q_{12}^{\a} }{(\boldsymbol{X}_{12}^{2})^{1/4}} \, ,
\end{align}
such that
\begin{align}
	\hat{\boldsymbol{X}}_{12}^{\a \s} \hat{\boldsymbol{X}}^{}_{21 \, \s \b} = \d_{\b}^{\a} \, , \hspace{10mm} \boldsymbol{J} = \hat{\Q}_{12}^{2} \, .
\end{align}
Now given an arbitrary three-point building block, $\boldsymbol{X}$, we construct the following higher-spin inversion operator:
\begin{equation}
	\cI_{\a(k) \b(k)}(\boldsymbol{X}) = \hat{\boldsymbol{X}}_{ (\a_{1} (\b_{1}} \dots \hat{\boldsymbol{X}}_{\a_{k}) \b_{k})}  \, . \label{Inversion tensor identities - three point functions a}
\end{equation}
This operators possess properties similar to the two-point higher-spin inversion operators \eqref{Higher-spin inversion operators a}. 
Let us now introduce the following analogues of the covariant spinor derivative and supercharge operators involving the three-point building blocks, 
where $(X,\Q) = (X_{12}, \Q_{12})$:
\begin{equation}
	\cD_{\a} = \frac{\partial}{\partial \Q^{\a}} + \text{i} (\g^{m})_{\a \b} \Q^{\b} \frac{\partial}{\partial X^{m}} \, , \hspace{5mm} \cQ_{ \a} = \text{i} \frac{\partial}{\partial \Q^{\a}} + (\g^{m})_{\a \b} \Q^{\b} \frac{\partial}{\partial X^{m}} \, , \label{Supercharge and spinor derivative analogues}
\end{equation}
which obey the commutation relations
\begin{equation}
	\big\{ \cD_{ \a} , \cD_{ \b} \big\} = \big\{ \cQ_{ \a} , \cQ_{ \b} \big\} = 2 \text{i} \, (\g^{m})_{\a \b} \frac{\partial}{\partial X^{m}} \, .
\end{equation}
Now given a function $f(\boldsymbol{X}_{12} , \Q_{12})$, there are the following differential identities which are essential for imposing 
differential constraints on three-point correlation functions of primary superfields:
\begin{subequations}
	\begin{align}
		D_{(1) \g} f(\boldsymbol{X}_{12} , \Q_{12}) &= (\boldsymbol{x}_{13}^{-1})_{\a \g} \cD^{\a} f(\boldsymbol{X}_{12} , \Q_{12}) \, ,  \label{Three-point building blocks 1c - differential identities 3} \\[2mm]
		D_{(2) \g} f(\boldsymbol{X}_{12} , \Q_{12}) &= \text{i} (\boldsymbol{x}_{23}^{-1})_{\a \g} \cQ^{\a} f(\boldsymbol{X}_{12} , \Q_{12}) \, .  \label{Three-point building blocks 1c - differential identities 4}
	\end{align}
\end{subequations}
Here by $D_{(1) \g}$ and $D_{(2) \g}$ we denote the ordinary superspace covariant derivatives acting on the superspace points $z_1= (x_1, \theta_1)$ and $z_2= (x_2, \theta_2)$
respectively. 

Now consider a primary tensor superfield $\F_{\cA}(z)$ of dimension $\D$ transforming in an irreducible representation of the Lorentz group. The two-point correlation function $\langle \F_{\cA}(z_{1}) \, \F^{\cB}(z_{2}) \rangle$ is constrained by superconformal symmetry to the following form:
\begin{equation} \label{Two-point correlation function}
	\langle \F_{\cA}(z_{1}) \, \F^{\cB}(z_{2}) \rangle = c \, \frac{\cI_{\cA}{}^{\cB}(\boldsymbol{x}_{12})}{(\boldsymbol{x}_{12}^{2})^{\D}} \, , 
\end{equation} 
where $\cI$ is an appropriate representation of the inversion tensor and $c$ is a constant real parameter. The denominator of the two-point function is determined by the conformal dimension of $\F_{\cA}$, which guarantees that the correlation function transforms with the appropriate weight under scale transformations.

For three-point functions, let $\F$, $\J$, $\P$ be primary superfields with scale dimensions $\D_{1}$, $\D_{2}$ and $\D_{3}$ respectively. The three-point function is
constructed using the general ansatz \cite{Park:1999pd,Park:1999cw}
\begin{align}
	\langle \F_{\cA_{1}}(z_{1}) \, \J_{\cA_{2}}(z_{2}) \, \P_{\cA_{3}}(z_{3}) \rangle = \frac{ \cI^{(1)}{}_{\cA_{1}}{}^{\cA'_{1}}(\boldsymbol{x}_{13}) \,  \cI^{(2)}{}_{\cA_{2}}{}^{\cA'_{2}}(\boldsymbol{x}_{23}) }{(\boldsymbol{x}_{13}^{2})^{\D_{1}} (\boldsymbol{x}_{23}^{2})^{\D_{2}}}
	\; \cH_{\cA'_{1} \cA'_{2} \cA_{3}}(\boldsymbol{X}_{12}, \Q_{12}) \, , \label{Three-point function - general ansatz}
\end{align} 
where the tensor $\cH_{\cA_{1} \cA_{2} \cA_{3}}$ encodes all information about the correlation function, and is related to the leading singular operator product expansion (OPE) coefficient \cite{Osborn:1993cr}. 

In this work we are primarily interested in the structure of three-point correlation functions of conserved (higher-spin) supercurrents. 
In 3D $\cN=1$ theories, a conserved (higher-spin) supercurrent of superspin-$s$ (integer or half-integer), is defined as a totally symmetric spin-tensor of rank $2s$, $\mathbf{J}_{\a_{1} \dots \a_{2s} }(z) = \mathbf{J}_{(\a_{1} \dots \a_{2s}) }(z) = \mathbf{J}_{\a(2s) }(z)$, satisfying a conservation equation of the form
\begin{equation} \label{Conserved current}
	D^{\a_{1}} \mathbf{J}_{\a_{1} \a_{2} \dots \a_{2s}}(z) = 0 \, .
\end{equation}
Conserved currents are primary superfields, and the dimension $\Delta_{\mathbf{J}}$ of $\mathbf{J}$ is fixed by the conservation condition \eqref{Conserved current} to $\D_{\mathbf{J}} = s+1$. At the component level, a higher-spin supercurrent of superspin-$s$ contains conserved conformal currents of spin-$s$ and spin-$(s+\tfrac{1}{2})$ respectively. Indeed, for conserved supercurrents of superspin $s$, the dimension $\D$ of the two-point function \eqref{Two-point correlation function} is fixed by conservation to $\D = s + 1$. If we now consider the three-point function of the conserved primary superfields $\mathbf{J}^{}_{\a(I)}$, $\mathbf{J}'_{\b(J)}$, $\mathbf{J}''_{\g(K)}$, where $I=2s_{1}$, $J=2s_{2}$, $K=2s_{3}$, then the general ansatz is
\begin{align} \label{Conserved correlator ansatz}
	\langle \, \mathbf{J}^{}_{\a(I)}(z_{1}) \, \mathbf{J}'_{\b(J)}(z_{2}) \, \mathbf{J}''_{\g(K)}(z_{3}) \rangle = \frac{ \cI_{\a(I)}{}^{\a'(I)}(\boldsymbol{x}_{13}) \,  \cI_{\b(J)}{}^{\b'(J)}(\boldsymbol{x}_{23}) }{(\boldsymbol{x}_{13}^{2})^{\D_{1}} (\boldsymbol{x}_{23}^{2})^{\D_{2}}}
	\; \cH_{\a'(I) \b'(J) \g(K)}(\boldsymbol{X}_{12}, \Q_{12}) \, ,
\end{align} 
where $\D_{i} = s_{i} + 1$. Below we summarise the constraints on $\cH$.
\begin{enumerate}
	\item[\textbf{(i)}] {\bf Homogeneity:}
	\begin{equation}
		\cH_{\a(I) \b(J) \g(K)}(\l^{2} \boldsymbol{X}, \l \Q) = (\l^{2})^{\D_{3} - \D_{2} - \D_{1}} \, \cH_{\a(I) \b(J) \g(K)}(\boldsymbol{X}, \Q) \, ,
	\end{equation}
	It is often convenient to introduce $\hat{\cH}_{\a(I) \b(J) \g(K)}(\boldsymbol{X}, \Q)$, such that
	\begin{align}
		\cH_{\a(I) \b(J) \g(K)}(\boldsymbol{X},\Q) &= \boldsymbol{X}^{\D_{3} - \D_{3}- \D_{1}} \hat{\cH}_{\a(I) \b(J) \g(K)}(\boldsymbol{X}, \Q) \, ,
	\end{align}
	where $\hat{\cH}_{\a(I) \b(J) \g(K)}(\boldsymbol{X}, \Q)$ is homogeneous degree 0 in $(\boldsymbol{X}, \Q)$, i.e.
	\begin{align}
		\hat{\cH}_{\a(I) \b(J) \g(K)}(\l^{2} \boldsymbol{X}, \l \Q) &= \hat{\cH}_{\a(I) \b(J) \g(K)}(\boldsymbol{X}, \Q) \, .
	\end{align}
	
	\item[\textbf{(ii)}] {\bf Differential constraints:} \\
	After application of the identities \eqref{Three-point building blocks 1c - differential identities 3}, \eqref{Three-point building blocks 1c - differential identities 4} we obtain the following constraints:
	\begin{subequations}
		\begin{align} 
			\text{Conservation at $z_{1}$:} && \cD^{\a} \cH_{\a \a(I - 1) \b(J) \g(K)}(\boldsymbol{X}, \Q) &= 0 \, , \label{Conservation equations on H - 1} \\
			\text{Conservation at $z_{2}$:} && \cQ^{\b} \cH_{\a(I) \b \b(J-1) \g(K)}(\boldsymbol{X}, \Q) &= 0 \, , \label{Conservation equations on H - 2} \\
			\text{Conservation at $z_{3}$:} && \cQ^{\g} \tilde{\cH}_{\a(I) \b(J) \g \g(K-1)  }(\boldsymbol{X}, \Q) &= 0 \label{Conservation equations on H - 3} \, ,
		\end{align}
	\end{subequations}
	where
	\begin{equation} \label{H tilde}
		\tilde{\cH}^{(\pm)}_{\a(I) \b(J) \g(K) }(\boldsymbol{X}, \Q) = (\boldsymbol{X}^{2})^{\D_{1} - \D_{3}} \, \cI_{\b(J)}{}^{\b'(J)}(\boldsymbol{X}) \, \cH^{I \, (\pm)}_{\a(I) \b'(J) \g(K)}(\boldsymbol{X}, \Q) \, . 
	\end{equation}

	\item[\textbf{(iii)}] {\bf Point-switch symmetries:} \\
	If the fields $\mathbf{J}$ and $\mathbf{J}'$ coincide, then we obtain the following point-switch identity
	\begin{equation} \label{Point-switch 1}
		\cH_{\a(I) \b(I) \g(K)}(\boldsymbol{X}, \Q) = (-1)^{\e(\mathbf{J})} \cH_{\b(I) \a(I) \g(K)}(-\boldsymbol{X}^{\text{T}}, -\Q) \, ,
	\end{equation}
	where $\e(\mathbf{J})$ is the Grassmann parity of $\mathbf{J}$. Likewise, if the fields $\mathbf{J}$ and $\mathbf{J}''$ coincide, then we obtain the constraint
	\begin{equation} \label{Point-switch 2}
		\tilde{\cH}_{\a(I) \b(J) \g(I) }(\boldsymbol{X}, \Q) = (-1)^{\e(\mathbf{J})} \cH_{\g(I) \b(J) \a(I)}(-\boldsymbol{X}^{\text{T}}, -\Q) \, .
	\end{equation}
\end{enumerate}

In the next sections, we will demonstrate that conservation on $z_{1}$ and $z_{2}$ is sufficient to constrain the structure of the three-point function to a unique parity-even solution, while the parity-odd solution must vanish.


\section{Grassmann-odd three-point functions -- \texorpdfstring{$\langle \mathbf{J}^{}_{F} \mathbf{J}'_{F} \mathbf{J}''_{F} \rangle$
	}{< F F F >}}\label{section3}

There are only two possibilities for Grassmann-odd three-point functions in superspace (up to permutations of the fields), they are:
\begin{equation}
	\langle \mathbf{J}^{}_{F} \mathbf{J}'_{F} \mathbf{J}''_{F} \rangle \, , \hspace{20mm} \langle \mathbf{J}^{}_{F} \mathbf{J}'_{B} \mathbf{J}''_{B} \rangle \, ,
\end{equation}
where ``$B$" represents a Grassmann-even (bosonic) field, and ``$F$" represents a Grassmann-odd (fermionic) field. Each of these correlation functions require separate analysis, however, they have a similar underlying structure.


\subsection{Method of irreducible decomposition}

First let us analyse the case $\langle \mathbf{J}^{}_{F} \mathbf{J}'_{F} \mathbf{J}''_{F} \rangle$; we consider three Grassmann-odd currents: 
$\mathbf{J}_{\a(2A+1)}$, $\mathbf{J}'_{\a(2B+1)}$, $\mathbf{J}''_{\g(2C+1)}$, where $A,B,C$ are positive integers. Therefore, the superfields $\mathbf{J}$, $\mathbf{J}'$, $\mathbf{J}''$ 
are of superspin 
$s_{1} = A+ \tfrac{1}{2}$, $s_{2} = B+ \tfrac{1}{2}$, $s_{3} = C+ \tfrac{1}{2}$ respectively. Using the formalism above, all information about the correlation function
\begin{equation}
	 \langle \, \mathbf{J}^{}_{\a(2A+1)}(z_{1}) \, \mathbf{J}'_{\b(2B+1)}(z_{2}) \, \mathbf{J}''_{\g(2C+1)}(z_{3}) \rangle \, ,
\end{equation}
is encoded in a homogeneous tensor field $\cH_{\a(2A+1) \b(2B+1) \g(2C+1)}(\boldsymbol{X}, \Q)$, which is a function of a single superspace 
variable $\cZ = ( \boldsymbol{X} , \Q)$ and satisfies the scaling property
\begin{equation}
	\cH_{\a(2A+1) \b(2B+1) \g(2C+1)}( \l^{2} \boldsymbol{X}, \l \Q) = (\l^{2})^{C-A-B-\tfrac{3}{2}}\cH_{\a(2A+1) \b(2B+1) \g(2C+1)}(\boldsymbol{X}, \Q) \, .
\end{equation}
To simplify the problem, for each set of totally symmetric spinor indices (the $\a$'s, $\b$'s and $\g$'s respectively), we convert pairs of them into vector indices as follows:
\begin{align}
	\cH_{\a(2A+1) \b(2B+1) \g(2C+1)}(\boldsymbol{X}, \Q) &\equiv \cH_{\a \a(2A), \b \b(2B), \g \g(2C)}(\boldsymbol{X}, \Q) \nonumber \\
	&= (\g^{m_{1}})_{\a_{1} \a_{2}} ... (\g^{m_{A}})_{\a_{2A-1} \a_{2A}} \nonumber \\
	& \times (\g^{n_{1}})_{\b_{1} \b_{2}} ... (\g^{n_{B}})_{\b_{2B-1} \b_{2B}} \nonumber \\
	& \times (\g^{k_{1}})_{\g_{1} \g_{2}} ... (\g^{k_{C}})_{\g_{2C-1} \g_{2C}} \nonumber \\
	& \times \cH_{\a \b \g, m_{1} ... m_{A}, n_{1} ... n_{B}, k_{1} ... k_{C} }(\boldsymbol{X}, \Q) \, .
\end{align}
The equality above holds only if and only if $\cH_{\a \b \g, m_{1} ... m_{A} n_{1} ... n_{B} k_{1} ... k_{C} }(\boldsymbol{X}, \Q)$ is totally symmetric
\begin{equation}
	\cH_{\a \b \g, m_{1} ... m_{A} n_{1} ... n_{B} k_{1} ... k_{C} }(\boldsymbol{X}, \Q) = \cH_{\a \b \g, (m_{1} ... m_{A}) (n_{1} ... n_{B}) (k_{1} ... k_{C}) }(\boldsymbol{X}, \Q) \, ,
\end{equation}
and traceless in each group of vector indices, i.e. $\forall i,j$
\begin{subequations} \label{FFF - traceless}
	\begin{align}
		\eta^{m_{i} m_{j}} \cH_{\a \b \g, m_{1} ... m_{i} m_{j} ... m_{A}, n_{1} ... n_{B}, k_{1} ... k_{C} }(\boldsymbol{X}, \Q) &= 0 \, ,
		\\
		\eta^{n_{i} n_{j}} \cH_{\a \b \g, m_{1} ... m_{A}, n_{1} ... n_{i} n_{j} ... n_{B}, k_{1} ... k_{C} }(\boldsymbol{X}, \Q) &= 0 \, , \\
		\eta^{k_{i} k_{j}} \cH_{\a \b \g, m_{1} ... m_{A}, n_{1} ... n_{B}, k_{1} ... k_{i} k_{j} ... k_{C} }(\boldsymbol{X}, \Q) &= 0 \, .
	\end{align}
\end{subequations}
It is also required that $\cH$ is subject to the $\g$-trace constraints
\begin{subequations} \label{FFF - Gamma trace}
	\begin{align}
		(\g^{m_{1}})_{\s}{}^{\a} \cH_{\a \b \g, m_{1} ... m_{A}, n_{1} ... n_{B}, k_{1} ... k_{C} }(\boldsymbol{X}, \Q) &= 0 \, , \label{FFF - Gamma trace 1} \\
		(\g^{n_{1}})_{\s}{}^{\b} \cH_{\a \b \g, m_{1} ... m_{A}, n_{1} ... n_{B}, k_{1} ... k_{C} }(\boldsymbol{X}, \Q) &= 0 \, , \label{FFF - Gamma trace 2} \\
		(\g^{k_{1}})_{\s}{}^{\g} \cH_{\a \b \g, m_{1} ... m_{A}, n_{1} ... n_{B}, k_{1} ... k_{C} }(\boldsymbol{X}, \Q) &= 0 \, . \label{FFF - Gamma trace 3}
	\end{align}
\end{subequations}
Now since $\cH$ is Grassmann-odd, it must be linear in $\Q$, and, using the property \eqref{Three-point building blocks 1a - properties 3}, we decompose $\cH$ as follows (raising all indices for convenience):
\begin{subequations} \label{FFF - irreducible decomposition}
\begin{equation}
	\cH^{\a \b \g, m(A) n(B) k(C) }(\boldsymbol{X}, \Q) = \sum_{i=1}^{4} \cH_{i}^{\a \b \g, m(A) n(B) k(C) }(X, \Q)\,,
\end{equation}
\vspace{-3mm}
\begin{align}
	\cH_{1}^{\a \b \g, m(A) n(B) k(C) }(X, \Q) &= \ve^{\a \b} \Q^{\g} A^{m(A) n(B) k(C)}(X) \, , \\
	\cH_{2}^{\a \b \g, m(A) n(B) k(C) }(X, \Q) &= \ve^{\a \b} (\gamma_{r})^{\g}{}_{\d} \Q^{\d} B^{r, m(A) n(B) k(C)}(X) \, , \\
	\cH_{3}^{\a \b \g, m(A) n(B) k(C) }(X, \Q) &= (\g_{p})^{\a \b} \Q^{\g} C^{p, m(A) n(B) k(C)}(X) \, , \\
	\cH_{4}^{\a \b \g, m(A) n(B) k(C) }(X, \Q) &= (\g_{p})^{\a \b} (\gamma_{r})^{\g}{}_{\d} \Q^{\d} D^{pr, m(A) n(B) k(C)}(X) \, .
\end{align}
\end{subequations}
Here it is convenient to view the contributions $\cH_{i}$ as functions of the symmetric tensor $X$ (which is equivalent to a three-dimensional vector) rather than of $\boldsymbol{X}$. 
In fact, since $\cH$ is linear in $\Q$ and $\Q^3=0$ we have $\cH (\boldsymbol{X}, \Q)= \cH (X, \Q)$. 
The tensors $A,B,C,D$ are constrained by conservation equations and any algebraic symmetry properties which $\cH$ possesses. In particular, the conservation equations \eqref{Conservation equations on H - 1}, \eqref{Conservation equations on H - 2}, are now equivalent to the following constraints on $\cH$ with vector indices
\begin{subequations}
	\begin{align} \label{FFF - conservation equations}
		\cD_{\a} \cH^{\a \b \g, m(A) n(B) k(C) }(\boldsymbol{X}, \Q) = 0 \, , \\
		\cQ_{\b} \cH^{\a \b \g, m(A) n(B) k(C) }(\boldsymbol{X}, \Q) = 0 \, .
	\end{align}
\end{subequations}
We also need consider the constraint for conservation at the third point \eqref{Conservation equations on H - 3}, however, this is technically challenging to impose using this analytic approach and we will not do it here. Instead we will comment on it at the end of subsections \ref{subsection3.3} and \ref{subsection3.4}. Since $\cH$ is linear in $\Q$, the conservation conditions \eqref{FFF - conservation equations} split up into constraints $O(\Q^{0})$
\begin{subequations}
	\begin{align} 
		\frac{\pa}{\pa \Q^{\a}} \cH^{\a \b \g, m(A) n(B) k(C) }(\boldsymbol{X}, \Q) &= 0  \, , \label{FFF - conservation equations O(0) - 1} \\
		\frac{\pa}{\pa \Q^{\b}} \cH^{\a \b \g, m(A) n(B) k(C) }(\boldsymbol{X}, \Q) &= 0 \, , \label{FFF - conservation equations O(0) - 2}
	\end{align}
\end{subequations}
and $O(\Q^{2})$
\begin{subequations}
	\begin{align} 
		(\g^{m})_{\a \d} \Q^{\d} \frac{\pa}{\pa X^{m}} \cH^{\a \b \g, m(A) n(B) k(C) }(\boldsymbol{X}, \Q) &= 0 \, , \label{FFF - conservation equations O(2) - 1} \\
		(\g^{m})_{\b \d} \Q^{\d} \frac{\pa}{\pa X^{m}} \cH^{\a \b \g, m(A) n(B) k(C) }(\boldsymbol{X}, \Q) &= 0 \, . \label{FFF - conservation equations O(2) - 2}
	\end{align}
\end{subequations}
Using the irreducible decomposition \eqref{FFF - irreducible decomposition}, equation \eqref{FFF - conservation equations O(0) - 1} results in the algebraic relations
\begin{subequations}
	\begin{align} \label{O(0) constraints - 1}
		A^{m(A) n(B) k(C)} + \eta_{pr} D^{pr, m(A) n(B) k(C) } &= 0 \, , \\
		B^{q,m(A) n(B) k(C)} + C^{q, m(A) n(B) k(C)} - \e^{q}{}_{pr} D^{pr, m(A) n(B) k(C)} &= 0 \, ,
	\end{align}
\end{subequations}
while \eqref{FFF - conservation equations O(0) - 2} gives
\begin{subequations}
	\begin{align} \label{O(0) constraints - 2}
		- A^{m(A) n(B) k(C)} + \eta_{pr} D^{pr, m(A) n(B) k(C) } &= 0 \, , \\
		- B^{q,m(A) n(B) k(C)} + C^{q, m(A) n(B) k(C)} - \e^{q}{}_{pr} D^{pr, m(A) n(B) k(C)} &= 0 \, .
	\end{align}
\end{subequations}
Hence, equations \eqref{O(0) constraints - 1}, \eqref{O(0) constraints - 2} together result in $A = B = 0$, while $C$ and $D$ satisfy:
\begin{subequations} \label{C and D constraints}
	\begin{align} 
		\eta_{pr} D^{pr, m(A) n(B) k(C) } &= 0 \, , \\
		C^{q, m(A) n(B) k(C)} - \e^{q}{}_{pr} D^{pr, m(A) n(B) k(C)} &= 0 \, .
	\end{align}
\end{subequations}
%
Next we consider the relations arising from the conservation equations at $O(\Q^{2})$. 
Using the decomposition \eqref{FFF - irreducible decomposition}, from a straightforward computation we obtain
\begin{subequations}
	\begin{align} \label{O(2) constraints}
		\pa_{t} \big( \e^{t}{}_{pr} D^{pr,m(A) n(B) k(C)} + C^{t,m(A) n(B) k(C)} \big) &= 0 \, , \\
		\pa_{t} \big( - \e^{tq}{}_{p} C^{p,m(A) n(B) k(C)} + D^{qt,m(A) n(B) k(C)} + D^{tq,m(A) n(B) k(C)}\big) &= 0 \, .
	\end{align}
\end{subequations}
After substituting the algebraic relations \eqref{C and D constraints} into \eqref{O(2) constraints}, we obtain
\begin{align} \label{C and D conservation equations}
	\pa_{p} C^{p,m(A) n(B) k(C)} = 0 \, , \hspace{10mm} \pa_{p} D^{pr, m(A) n(B) k(C)} = 0 \, .
\end{align}
We now must impose the $\g$-trace conditions, starting with \eqref{FFF - Gamma trace 1} and \eqref{FFF - Gamma trace 2}. Making use of the decomposition \eqref{FFF - irreducible decomposition}, equation \eqref{FFF - Gamma trace 1} results in the algebraic constraints
\begin{subequations}
	\begin{align}
		\eta_{mp} C^{p, m m(A-1) n(B) k(C)} &= 0 \, , & \e_{qmp} C^{p, m m(A-1) n(B) k(C)} &= 0 \, , \\
		\eta_{mp} D^{pr, m m(A-1) n(B) k(C)} &= 0 \, , & \e_{qmp} D^{pr, m m(A-1) n(B) k(C)} &= 0 \, , 
	\end{align}
\end{subequations}
while from \eqref{FFF - Gamma trace 2} we find
\begin{subequations}
	\begin{align}
		\eta_{np} C^{p, m(A) n n(B-1) k(C)} &= 0 \, , & \e_{qnp} C^{p, m(A) n n(B-1) k(C)} &= 0 \, , \\
		\eta_{np} D^{pr, m(A) n n(B-1) k(C)} &= 0 \, , & \e_{qnp} D^{pr, m(A) n n(B-1) k(C)} &= 0 \, .
	\end{align}
\end{subequations}
Altogether these relations imply that both $C$ and $D$ are symmetric and traceless in the indices $p, m_{1}, ..., m_{A}, n_{1}, ..., n_{B}$, i.e.,
\begin{equation}
	C^{p, m(A) n(B) k(C)} \equiv C^{(p m(A) n(B)) k(C)} \, , \hspace{5mm} D^{pr, m(A) n(B) k(C)} \equiv D^{(p m(A) n(B)), r k(C)} \, .
\end{equation}
Next, from the $\g_{k}$-trace constraint \eqref{FFF - Gamma trace 3}, we obtain the algebraic relations
\begin{subequations}  \label{C and D constraints 2} 
	\begin{align}
		\eta_{kr} D^{pr, m(A) n(B) k k(C-1) } = 0 \, , \\
		C^{p, m(A) n(B) k k(C-1)} + \e^{k}{}_{rs} D^{pr, m(A) n(B) s k(C-1)} = 0 \, .
	\end{align}
\end{subequations}
To make use of these relations, we decompose $D$ into symmetric and anti-symmetric parts on the indices $r, k_{1}, ..., k_{C}$ as follows:
\begin{align} \label{D decomposition}
	D^{(p m(A) n(B)), r ( k_{1} ... k_{C} ) } = D_{S}^{ (p m(A) n(B)), ( r k_{1} ... k_{C} ) } + \sum_{i=1}^{C} \e^{r k_{i}}{}_{t} D_{A}^{ (p m(A) n(B)), ( t k_{1} ... \hat{k}_{i} ... k_{C} ) } \, ,
\end{align}
where 
the notation $\hat{k}_{i}$ denotes removal of the index $k_{i}$ from $D_{A}$. After substituting this decomposition into \eqref{C and D constraints 2}, we obtain:
\begin{subequations}
	\begin{equation}
		\eta_{k r}  D_{S}^{ (p m(A) n(B)), ( r k k(C-1) ) } = 0 \, ,
	\end{equation}
	\begin{equation}
		D_{A}^{  (p m(A) n(B)), ( r k(C-1) ) } = \frac{1}{C+1} C^{ (p m(A) n(B)), ( r k(C-1) ) } \, .
	\end{equation}
\end{subequations}
We see that the tensor $D_A$ is fully determined in terms of $C$. 
To continue we now substitute \eqref{D decomposition} into \eqref{C and D constraints} to obtain equations relating $C$ and $D_{S}$; we obtain:
\begin{subequations} \label{C and D algebraic relations}
	\begin{align}
		&\eta_{pr} D_{S}^{(p m(A) n(B)),(r k(C))} + \frac{1}{C+1} \sum_{i=1}^{C} \e^{k_{i}}{}_{p r} C^{ (p m(A) n(B)), ( r k_{1} ... \hat{k}_{i} ... k_{C} ) } = 0 \, , \\
		&\e^{p}{}_{qr} D_{S}^{(q m(A) n(B)), (r k(C))} - C^{(p m(A) n(B)), (r k(C))} \\ 
		&+ \frac{1}{C+1} \sum_{i=1}^{C} \Big\{ \eta^{p k_{i}} \eta_{qr} C^{(q m(A) n(B)), (r k_{1} ... \hat{k}_{i} ... k(C)) } - C^{(k_{i} m(A) n(B)),(p k_{1} ... \hat{k}_{i} ... k(C))} \Big\} = 0 \, . \nonumber
	\end{align}
\end{subequations}
Further, the conservation equations \eqref{C and D conservation equations} are now equivalent to
\begin{equation}
		\pa_{p} C^{(pm(A) n(B)), k(C)} = 0 \, , \hspace{10mm} \pa_{p} D_{S}^{(p m(A) n(B)),(rk(C))} = 0 \, .
		\label{e4}
\end{equation}
Hence, finding the solution for the three-point function $\langle \mathbf{J}^{}_{F} \mathbf{J}'_{F} \mathbf{J}''_{F} \rangle$ is now equivalent to finding two transverse tensors $C$ and $D_{S}$, which are related by the algebraic constraints \eqref{C and D algebraic relations}. It may be checked for $A=B=C=1$ that the constraints reproduce those of the supercurrent three-point 
function found in~\cite{Buchbinder:2015qsa}. 

Let us now briefly comment on the analysis of three-point functions involving flavour currents, i.e. when $A,B,C = 0$. In these cases one can simply ignore the relevant tracelessness and $\g$-trace conditions \eqref{FFF - traceless}, \eqref{FFF - Gamma trace} respectively and omit the appropriate groups of tensor indices. The analysis of the conservation equations proves to be essentially the same and we will not elaborate further on these cases.


\subsection{Conservation equations}

Let us now summarise the constraint analysis in the previous subsection in a way that makes the symmetries more apparent. For the $\langle \mathbf{J}^{}_{F} \mathbf{J}'_{F} \mathbf{J}''_{F} \rangle$ correlators we have two tensors; $C$ of rank $N+C+1$, and $D$ of rank $N+C+2$, where $N= A +B$, which possess the following symmetries:
\begin{align}
	C^{(r_{1} ... r_{N+1})(k_{1} ... k_{C})} \, , \hspace{10mm} D_{S}^{(r_{1} ... r_{N+1})(k_{1} ... k_{C+1})} \, .
\end{align}
The tensors $C$ and $D_{S}$ are totally symmetric and traceless in the groups of indices $r$ and $k$ respectively. They also satisfy the conservation conditions
\begin{align} \label{C and D conservation equations - more symmetric}
	\pa_{r_{1}} C^{(r_{1} ... r_{N+1})(k_{1} ... k_{C})} = 0 \, , \hspace{5mm} \pa_{r_{1}} D_{S}^{(r_{1} ... r_{N+1})(k_{1} ... k_{C+1})} = 0 \, ,
\end{align}
and the algebraic relations
\begin{subequations} \label{C and D algebraic relations - more symmetric}
	\begin{align}
		&\eta_{r_{1} k_{1}} D_{S}^{(r_{1} ... r_{N+1})(k_{1} ... k_{C+1})} + \frac{1}{C+1} \sum_{i=2}^{C+1} \e^{k_{i}}{}_{r_{1} k_{1}} C^{ (r_{1} ... r_{N+1})( k_{1} k_{2} ... \hat{k}_{i} ... k_{C+1} ) } = 0 \, , \\
		&\e^{p}{}_{r_{1} k_{1}} D_{S}^{(r_{1} ... r_{N+1})(k_{1} ... k_{C+1})} - C^{(p r_{2} ... r_{N+1})(k_{2} ... k_{C+1})} \\ 
		&+ \frac{1}{C+1} \sum_{i=2}^{C+1} \Big\{ \eta^{p k_{i}} \eta_{r_{1} k_{1}} C^{(r_{1} ... r_{N+1})(k_{1} k_{2} ... \hat{k}_{i} ... k_{C+1}) } - C^{(k_{i} r_{2} ... r_{N+1})(p k_{2} ... \hat{k}_{i} ... k_{C+1})} \Big\} = 0 \, . \nonumber
	\end{align}
\end{subequations}
The ``full" tensor $D$, present in the decomposition \eqref{FFF - irreducible decomposition}, is constructed from $C$ and $D_{S}$ as follows:
\begin{align} \label{D decomposition - more symmetric}
	D^{q, (r_{1} ... r_{N+1})( k_{1} ... k_{C} ) } = D_{S}^{ (r_{1} ... r_{N+1})( q k_{1} ... k_{C} ) } + \frac{1}{C+1} \sum_{i=1}^{C} \e^{q k_{i}}{}_{t} C^{ (r_{1} ... r_{N+1})( t k_{1} ... \hat{k}_{i} ... k_{C} ) } \, .
\end{align}
As we will see later, the algebraic relations \eqref{C and D algebraic relations - more symmetric} are sufficient to determine $D$ completely in terms of $C$. 
However, before we prove this it is prudent to analyse the conservation equation \eqref{C and D conservation equations - more symmetric} for $C$.

Since we have identified the algebraic symmetries of $C$, it is convenient to convert $C$ back into spinor notation and contract it with commuting auxiliary spinors as follows:
\begin{align}
	C(X; u(2N+2), v(2C)) = C_{\a(2N+2) \b(2C)}(X) \, u^{\a_{1}} \, ... \, u^{\a_{2N+2}} v^{\b_{1}} \, ... \, v^{\b_{2C}} \, .
\end{align}
Since the auxiliary spinors are commuting they satisfy
\begin{align}
	\ve_{\a \b} u^{\a} u^{\b} = 0 \, , \hspace{15mm}  \ve_{\a \b} v^{\a} v^{\b} = 0 \, .
\end{align}
We now introduce a basis of monomials out of which $C$ can be constructed. Adapting the results of \cite{Buchbinder:2022mys,Buchbinder:2023fqv}, we use
\begin{align} \label{Basis structures}
	P_{3} &= \ve_{\a \b} u^{\a} v^{\b} \, , & Q_{3} &= \hat{X}_{\a \b} u^{\a} v^{\b} \, ,  & Z_{1} &= \hat{X}_{\a \b} u^{\a} u^{\b} \, , & Z_{2} &= \hat{X}_{\a \b} v^{\a} v^{\b} \, .
\end{align}
A general ansatz for $C(X;u,v)$ which is homogeneous degree $2(N+1)$ in $u$, $2C$ in $v$ and $C-N-2$ in $X$ is of the following form:
\begin{equation}
	C(X; u,v ) = X^{C-N-2} \sum_{a,b} \a(a,b) P_{3}^{a} Q_{3}^{2(C-b) - a} Z_{1}^{b + N - C + 1} Z_{2}^{b} \, .
\end{equation}
However, there is linear dependence between the monomials \eqref{Basis structures} of the form
\begin{equation}
	Z_{1} Z_{2} = Q_{3}^{2} - P_{3}^{2} \, ,
\end{equation}
which allows for elimination of $Z_{2}$. Hence, the ansatz becomes:
\begin{equation} \label{C polynomial ansatz}
	C(X; u,v ) = X^{C-N-2} \sum_{k=0}^{2C} \a_{k} P_{3}^{k} Q_{3}^{2C - k} Z_{1}^{N-C+1} \, .
\end{equation}
This expansion is valid for $C \leq N +1$, which
is always fulfilled for field configurations where the third superspin triangle inequality, $s_{3} \leq s_{1} + s_{2} $, is satisfied. 
Note that if $s_{3} > s_{1} + s_{2} $ it then follows that $s_{1} \leq s_{2} + s_{3}$ and $s_{2} \leq s_{1} + s_{3}$. That is, if one of the triangle inequalities is 
violated the remaining two are necessarily satisfied. It implies that one can always arrange the fields in the three-point function to obtain a configuration for which $s_{3} \leq s_{1} + s_{2} $, 
or equivalently, 
$C \leq N < N +1$. We will assume that we have performed such an arrangement and use eq.~\eqref{C polynomial ansatz}. 

Requiring that the three-point function is conserved at $z_{1}$ and $z_{2}$ is now tantamount to imposing
\begin{equation}
	\frac{\pa}{\pa u^{\a}} \frac{\pa}{\pa u^{\b}} \frac{\pa}{\pa X_{ \a \b}} C(X; u,v ) = 0 \, .
\end{equation}
By acting with this operator on the ansatz \eqref{C polynomial ansatz}, we obtain
\begin{align}
	X^{C-N-3} \sum_{k=0}^{2C} \a_{k} P_{3}^{k-2} Q_{3}^{2C - 2 - k} Z_{1}^{N-C} \Big\{ \s_{1}(k) P_{3}^{2} Q_{3}^{2} + \s_{2}(k) Q_{3}^{4} + \s_{3}(k) P_{3}^{4} \Big\} = 0 \, ,
\end{align}
where
\begin{subequations}
	\begin{align}
		\s_{1}(k) &= - 2 k^3 + 4 C k^2 + 2 k (1 + C + 2 N (2 + N)) - 2 C (3 + 4 N (2 + N)) \, , \\
		\s_{2}(k) &= - k(k-1) (2 N - k + 3)  \, , \\
		\s_{3}(k) &= (2 C - k - 1) (2 C - k) (2 N + k + 3) \, .
	\end{align}
\end{subequations}
The sum above may now be split up into three contributions so that the coefficients may be easily read off
\begin{align}
	\sum_{k=0}^{2C} \a_{k} \s_{1}(k) P_{3}^{k} Q_{3}^{2C - k} + \sum_{k=0}^{2C-2} \a_{k+2} \s_{2}(k+2) P_{3}^{k} Q_{3}^{2C - k} + \sum_{k=2}^{2C} \a_{k-2} \s_{3}(k-2) P_{3}^{k} Q_{3}^{2C - k} = 0\, .
\end{align}
Hence, we obtain the following linear system:
\begin{subequations} \label{Linear homogeneous equations}
	\begin{align}
		\a_{k} \s_{1}(k) + \a_{k+2} \s_{2}(k+2) + \a_{k-2} \s_{3}(k-2) = 0 \, , \hspace{10mm} 2 \leq k \leq 2C-2 \, ,
	\end{align}
	\vspace{-12mm}
	\begin{align}
		\a_{0} \s_{1}(0) + \a_{2} \s_{2}(2) &= 0 \, , & \a_{2C} \s_{1}(2C) + \a_{2C-2} \s_{3}(2C-2) &= 0 \, , \\[1mm]
		\a_{1} \s_{1}(1) + \a_{3} \s_{2}(3) &= 0 \, , & \a_{2C-1} \s_{1}(2C-1) + \a_{2C-3} \s_{3}(2C-3) &= 0 \, .
	\end{align}
\end{subequations}
It must be noted that the equations above for the variables $\a_{k}$ split into independent $C+1$ and $C$ dimensional systems of linear homogeneous equations corresponding to the parity-even and parity-odd sectors respectively. The terms for which $k$ is even are denoted parity-even, while the terms for which $k$ is odd are denoted parity-odd, so that
\begin{equation}
	C(X;u,v) = C_{E}(X;u,v) + C_{O}(X;u,v) \, ,
\end{equation}
where
\begin{subequations}
	\begin{align}
		C_{E}(X; u,v ) &= X^{C-N-2} \sum_{k=0}^{C} \a_{2k} P_{3}^{2k} Q_{3}^{2(C - k)} Z_{1}^{N-C+1} \, , \label{zh7} \\
		C_{O}(X; u,v ) &= X^{C-N-2} \sum_{k=1}^{C} \a_{2k-1} P_{3}^{2k-1} Q_{3}^{2(C - k) + 1} Z_{1}^{N-C+1} \, .
	\end{align}
\end{subequations}

\noindent
Indeed, this convention is consistent with that of \cite{Buchbinder:2022mys,Buchbinder:2023fqv}. Hence, in the linear homogeneous system \eqref{Linear homogeneous equations}, we define the parity-even coefficients, $a_{k}$, $b_{k}$, $c_{k}$, as
\begin{subequations}
	\begin{align}
		a_{k} &= \s_{1}(2k-2) \, , & &1\leq k \leq C+1 \, ,\\
		b_{k} &= \s_{2}(2k) \, , & &1\leq k \leq C \, , \\
		c_{k} &= \s_{3}(2k-2) \, , & &1\leq k \leq C \, ,
	\end{align}
\end{subequations}
and the parity-odd coefficients, $\tilde{a}_{k}$, $\tilde{b}_{k}$, $\tilde{c}_{k}$ as
\begin{subequations}
	\begin{align}
		\tilde{a}_{k} &= \s_{1}(2k-1) \, , & &1\leq k \leq C \, , \\
		\tilde{b}_{k} &= \s_{2}(2k+1) \, , & &1\leq k \leq C-1 \, , \\
		\tilde{c}_{k} &= \s_{3}(2k-1) \, , & &1\leq k \leq C-1 \, .
	\end{align}
\end{subequations}
With the above definitions, the linear homogeneous equations \eqref{Linear homogeneous equations} split into two independent systems which can be written in the form 
$\mathbf{M} \vec{\a} = \mathbf{0}$. More explicitly
\begin{subequations} \label{zh100}
	\begin{align}
		\textbf{Even:}& &
		\mathbf{M}_{E} &= 
		\setlength{\arraycolsep}{5pt}
		\begin{bmatrix}
			\; a_{1} & b_{1} & 0 & 0 & \dots & 0 \\
			\; c_{1} & a_{2} & b_{2} & 0 & \dots & 0 \\
			\; 0 & c_{2} & a_{3} & b_{3} & \dots & 0 \\
			\; \vdots & \vdots & \ddots & \ddots & \ddots & \vdots \\
			\; 0 & \dots & \dots & c_{C-1} & a_{C} & b_{C} \\
			\; 0 & 0 & \dots & 0 & c_{C} & a_{C+1}
		\end{bmatrix} , &
		\vec{\a}_{E} &=
		\begin{bmatrix}
			\a_{0} \\
			\a_{2} \\
			\a_{4} \\
			\vdots \\
			\a_{2C-2} \\
			\a_{2C} 
		\end{bmatrix} , \label{Parity even homogeneous system}\\[5mm]
		\textbf{Odd:}& &
		\mathbf{M}_{O} &= 
		\setlength{\arraycolsep}{5pt}
		\begin{bmatrix}
			\; \tilde{a}_{1} & \tilde{b}_{1} & 0 & 0 & \dots & 0 \\
			\; \tilde{c}_{1} & \tilde{a}_{2} & \tilde{b}_{2} & 0 & \dots & 0 \\
			\; 0 & \tilde{c}_{2} & \tilde{a}_{3} & \tilde{b}_{3} & \dots & 0 \\
			\; \vdots & \vdots & \ddots & \ddots & \ddots & \vdots \\
			\; 0 & \dots & \dots & \tilde{c}_{C-2} & \tilde{a}_{C-1} & \tilde{b}_{C-1} \\
			\; 0 & 0 & \dots & 0 & \tilde{c}_{C-1} & \tilde{a}_{C}
		\end{bmatrix} , &
		\vec{\a}_{O} &=
		\begin{bmatrix}
			\a_{1} \\
			\a_{3} \\
			\a_{5} \\
			\vdots \\
			\a_{2C-3} \\
			\a_{2C-1} 
		\end{bmatrix} , \label{Parity odd homogeneous system}
	\end{align}
\end{subequations}
where $\mathbf{M}_{E}$, $\mathbf{M}_{O}$ are square matrices of dimension $C+1$, $C$ respectively. The system of equations \eqref{Parity even homogeneous system} is associated with the solution $C_{E}(X;u,v)$, while the system \eqref{Parity odd homogeneous system} is associated with the solution $C_{O}(X;u,v)$. The question now is whether there exists explicit solutions to these linear homogeneous systems for arbitrary $A,B,C$. 

The matrices of the form~\eqref{zh100} are referred to as tri-diagonal matrices,
before we continue with the analysis let us comment on some of their features.
The sufficient conditions under which a tri-diagonal matrix is invertible for arbitrary sequences $a_{k}, b_{k}, c_{k}$ has been discussed in e.g. \cite{BRUGNANO1992131,YHuang_1997}. 
Now consider the determinant of a tri-diagonal matrix
\begin{align}
	\D_{k} = 
	\setlength{\arraycolsep}{5pt}
	\begin{vmatrix}
		\; a_{1} & b_{1} & 0 & 0 & \dots & 0 \\
		\; c_{1} & a_{2} & b_{2} & 0 & \dots & 0 \\
		\; 0 & c_{2} & a_{3} & b_{3} & \dots & 0 \\
		\; \vdots & \vdots & \ddots & \ddots & \ddots & \vdots \\
		\; 0 & \dots & \dots & c_{k-2} & a_{k-1} & b_{k-1} \; \\
		\; 0 & 0 & \dots & 0 & c_{k-1} & a_{k}
	\end{vmatrix} \, .
\end{align}
One of its most important properties is that it satisfies the recurrence relation
\begin{equation} \label{Continuant equation}
	\D_{k} = a_{k} \D_{k-1} - b_{k-1} c_{k-1} \D_{k-2} \, , \hspace{10mm} \D_{0} = 1 \, , \hspace{4mm} \D_{-1} = 0 \, ,
\end{equation}
which may be seen by performing a Laplace expansion on the last row. The sequence $\D_{k}$ is often called the ``generalised continuant" with respect to the sequences $a_{k}, b_{k}, c_{k}$. There are methods to compute this determinant in closed form for simple cases, for example if the tri-diagonal matrix under consideration is ``Toeplitz" 
(i.e. if the sequences are constant $a_{k} = a$, $b_{k} = b$, $c_{k} = c$). 
For Toeplitz tri-diagonal matrices one obtains
%
\begin{align}
	\D_{k} = \frac{1}{\sqrt{a^{2} - 4 b c}} \, \Bigg\{ \bigg(\frac{a+\sqrt{a^{2} - 4 b c}}{2} \bigg)^{k+1} - \: \bigg(\frac{a - \sqrt{a^{2} - 4 b c}}{2}\bigg)^{k+1} \Bigg\} \, ,
\end{align}
for $a^{2} - 4 b c \neq 0$, 
while for $a^{2} - 4 b c = 0$ we obtain $\D_{k} = (k+1) (\frac{a}{2})^{k}$. For general sequences $a_{k}, b_{k}, c_{k}$ there is no straightforward approach to compute $\D_{k}$ and it must computed recursively using \eqref{Continuant equation}. Another important feature of tri-diagonal matrices is that in general their nullity (co-rank) is either $0$ or $1$,
which implies that any system of linear homogeneous equations with a tri-diagonal matrix has at most one non-trivial solution.

In the next subsections, we study the continuants of $\mathbf{M}_{E},\mathbf{M}_{O}$ for the homogeneous systems \eqref{Parity even homogeneous system}, \eqref{Parity odd homogeneous system}, and obtain their explicit form for arbitrary $A,B,C$. This determines whether $\mathbf{M}_{E},\mathbf{M}_{O}$ are invertible, and the number of solutions for the homogeneous systems \eqref{Parity even homogeneous system}, \eqref{Parity odd homogeneous system}. 


\subsection{Parity-odd case}\label{subsection3.3}

First we will analyse the system of equations for the parity-odd sector. 
Note that if $\det[ \mathbf{M}_{O} ] \neq 0$ then the system of equations 
$\mathbf{M}_{O} \vec{\a}_{O} = \mathbf{0}$ admits only the trivial solution. 
Let us denote $\mathbf{M}_{O} := \mathbf{M}_{O}^{(C)}$ to indicate that the dimension of the tri-diagonal matrix in~\eqref{Parity odd homogeneous system} is $C \times C$.
We also introduce the $ k \times k$ continuant $\tilde{\D}_{k}^{(C)}$, where $1  \leq k \leq C$. It satisfies the continuant equation~\eqref{Continuant equation} with
\begin{subequations} \label{Parity odd coefficients}
	\begin{align}
		\tilde{a}_{k} &= 4 C (-1 + k (4k-3) - 2 N (N+2) ) \nonumber\\
		& \hspace{10mm} - 4 (2 k - 1) (2 k (k - 1) - N (N + 2))  \, , & &1\leq k \leq C \, , \\
		\tilde{b}_{k} &= - 4 k (1 + 2 k) (N - k+1) \, , & &1\leq k \leq C-1 \, , \\
		\tilde{c}_{k} &= 4 (1 + 2 C - 2 k) (C - k) (1 + k + N) \, , & &1\leq k \leq C-1 \, .
	\end{align}
\end{subequations}
We also have $\tilde{\D}_{C}^{(C)} = \det[ \mathbf{M}_{O}^{(C)} ]$. Below we present some examples of the matrix $\mathbf{M}_{O}$ and the determinant 
$\tilde{\D}_{C}^{(C)}$ for arbitrary $N=A+B$ and fixed $C$:
\begin{subequations} \label{Parity odd matrix examples}
	\begin{align}
		\mathbf{M}_{O}^{(1)} = \left[ \,
			-4 N (N+2) \, \right] \, ,
	\end{align}
\begin{align}
	\mathbf{M}_{O}^{(2)} = 
	\left[
	\begin{array}{cc}
		-12 N (N+2) & -12 N \\
		12 (N+2) & 24-4 N (N+2) \\
	\end{array}
	\right] ,
\end{align}
\begin{align}
	\mathbf{M}_{O}^{(3)} = 
	\left[
	\begin{array}{ccc}
		-20 N (N+2) & -12 N & 0 \\
		40 (N+2) & 60-12 N (N+2) & -40 (N-1) \\
		0 & 12 (N+3) & 72-4 N (N+2) \\
	\end{array}
	\right] , 
\end{align}
\begin{align}
	\mathbf{M}_{O}^{(4)} = 
	\left[
	\scalemath{0.9}{\begin{array}{cccc}
		-28 N (N+2) & -12 N & 0 & 0 \\
		84 (N+2) & 96-20 N (N+2) & -40 (N-1) & 0 \\
		0 & 40 (N+3) & 176-12 N (N+2) & -84 (N-2) \\
		0 & 0 & 12 (N+4) & 144-4 N (N+2) \\
	\end{array}}
	\right] \,.
\end{align}
\end{subequations}
The corresponding determinants are:
\begin{subequations}
	\begin{align}
		\tilde{\D}_{1}^{(1)} &= -4 N (N+2) \, ,\\
		\tilde{\D}_{2}^{(2)} &= 48 (N-1) N (N+2) (N+3) \, , \\
		\tilde{\D}_{3}^{(3)} &= -960 (N-2) (N-1) N (N+2) (N+3) (N+4) \, , \\
		\tilde{\D}_{4}^{(4)} &= 26880 (N-3) (N-2) (N-1) N (N+2) (N+3) (N+4) (N+5) \, . 
	\end{align}
\end{subequations}
Indeed the matrices above are invertible as $\det[\mathbf{M}_{O}] \neq 0$, therefore we have the solution $\vec{\a}_{O} = \mathbf{0}$. The determinant can be efficiently computed using the recursion formula \eqref{Continuant equation}, and the pattern appears holds for large integers $N,C$. Using Mathematica we explicitly computed $\tilde{\D}_{C}^{(C)}$ for arbitrary $N$, up to $C=500$. In all cases we found it is non-trivial.

However, the continuant $\tilde{\D}_{k}^{(C)}$ can also be obtained explicitly for all $1 \leq k \leq C$, and arbitrary $C$. First, one can show that $\tilde{\D}_{k}^{(C)}$ satisfies the following recurrence relation:
\begin{align} \label{Trick recursion - odd sector}
	\tilde{\D}_{k}^{(C)} = \tilde{\g}_{k-1}^{(C)} \tilde{\D}_{k-1}^{(C)} \, , \hspace{5mm} \tilde{\D}_{1}^{(C)} = -4(2C-1) N (2+N) \, , \hspace{5mm} 1 \leq k \leq C \, ,
\end{align}
where $\tilde{\g}^{(C)}_{k}$ is given by 
%
\begin{align} \label{e1-Odd}
	\tilde{\g}^{(C)}_{k} = - 4 (N-k) (2 + k + N ) (-1 + 2C - 2k ) \, .
\end{align}
To see this, consider the combination $\tilde{\D}_{k+2}^{(C)} - \tilde{a}_{k+2} \tilde{\D}_{k+1}^{(C)} + \tilde{b}_{k+1} \tilde{c}_{k+1} \tilde{\D}_{k}^{(C)}$. 
Using eqs. \eqref{Trick recursion - even sector}, \eqref{e1-Odd}, this combination becomes $( \tilde{\g}_{k+1} \tilde{\g}_{k} - \tilde{a}_{k+2} \tilde{\g}_{k}  + \tilde{b}_{k+1} \tilde{c}_{k+1} ) \tilde{\D}^{(C)}_{k}$. 
However it may be shown using eqs.~\eqref{Parity odd coefficients} and~\eqref{e1-Odd} that 
\begin{equation}
	\label{e2}
	\tilde{\g}_{k+1} \tilde{\g}_{k} - \tilde{a}_{k+2} \tilde{\g}_{k}  + \tilde{b}_{k+1} \tilde{c}_{k+1} = 0 \, ,
\end{equation}
for arbitrary $k, N, C$, which implies that the recurrence relation~\eqref{Continuant equation} is indeed satisfied. We find the following general solution for \eqref{Trick recursion - odd sector}:\footnote{Recall that we have arranged the operators in the three-point function so that $s_3 \leq s_1+ s_2$ which implies $C \leq N$.}
\begin{align} \label{Parity-odd continuant}
	\tilde{\D}_{k}^{(C)} = 2^{3k} \, \frac{\Gamma(\tfrac{1}{2} - C + k) }{\Gamma(\tfrac{1}{2}-C) } \frac{(N+k+1)!}{ (N+1) (N-k)!} \, , \hspace{5mm} 1 \leq k \leq C \, .
\end{align}
One can check (using e.g. Mathematica) that it solves the continuant equation~\eqref{Continuant equation}. Recalling that $\det[ \mathbf{M}_{O}^{(C)} ] = \tilde{\D}_{C}^{(C)}$, it then follows that 
\begin{align} \label{Odd determinant}
\det[ \mathbf{M}_{O}^{(C)} ] = \frac{2^{3C} \sqrt{\pi }}{\Gamma (\tfrac{1}{2}-C)} \frac{ (N+C+1)!}{ (N+1) (N-C)!} \, , \hspace{5mm} C \leq N \, ,
\end{align}
%
which is always non-trivial. This implies that $\mathbf{M}_{O}$ is of full rank and, hence, the system of equations \eqref{Parity odd homogeneous system} admits only the trivial solution $\vec{\a}_{O} = \mathbf{0}$. 
Therefore, recalling that $C_{O}$ is the parity-odd solution corresponding to the system of equations \eqref{Parity odd homogeneous system}, from the analysis above we have shown 
that $C_{O}(X;u,v) = 0$ for arbitrary $A, B, C$. 

Now we will show that the tensor $D_{S}$ associated with $C_{O}$, which are related by the algebraic relations \eqref{C and D algebraic relations - more symmetric}, also vanishes. 
Since we have shown that $C_{O}=0$ in general, eq.~\eqref{C and D algebraic relations - more symmetric} implies
\begin{subequations}
	\begin{align}
		\eta_{r_{1} k_{1}} D_{S}^{(r_{1} ... r_{N+1})(k_{1} ... k_{C+1})} &= 0 \, , \\
		\e^{p}{}_{r_{1} k_{1}} D_{S}^{(r_{1} ... r_{N+1})(k_{1} ... k_{C+1})} &= 0 \, .
	\end{align}
\end{subequations}
Hence, $D_{S}$ is totally symmetric and traceless in all tensor indices
\begin{equation}
	 D_{S}^{(r_{1} ... r_{N+1})(k_{1} ... k_{C+1})} \equiv D_{S}^{(r_{1} ... r_{N+1} k_{1} ... k_{C+1})} \, .
\end{equation}
We can now construct a solution for $D_{S}$ using auxiliary spinors as follows:
\begin{align}
	D_{S}(X; u(2N+2C+4) ) = D_{S \, \a(2N+2C+4)}(X) \, u^{\a_{1}} \, ... \, u^{\a_{2N+2C+4}} \, .
\end{align}
Since $D_{S}$ is transverse, the polynomial $D_{S}(X;u)$ must satisfy the conservation equation
\begin{equation} \label{D conservation equation}
	\frac{\pa}{\pa u^{\a}} \frac{\pa}{\pa u^{\b}} \frac{\pa}{\pa X_{ \a \b}} D_{S}(X; u) = 0 \, .
\end{equation}
The only possible structure (up to a constant coefficient) for $D_{S}(X;u)$ is of the form
\begin{equation}
	D_{S}(X; u(2N+2C+4) ) =X^{C-N-2} Z_{1}^{N+C+2} \, .
\end{equation}
Explicit computation of \eqref{D conservation equation} gives
\begin{equation}
	\frac{\pa}{\pa u^{\a}} \frac{\pa}{\pa u^{\b}} \frac{\pa}{\pa X_{ \a \b}} D_{S}(X; u) = -4(N+C+2)^{2}(1+2C) \, X^{C-N-3} Z_{1}^{N+C+1} \, ,
\end{equation}
which is always non-zero. Therefore $D_{S} = 0$ for the parity-odd sector. Hence, for Grassmann-odd three-point functions of the 
form $\langle \mathbf{J}^{}_{F} \mathbf{J}'_{F} \mathbf{J}''_{F} \rangle$, there is no parity-odd solution for arbitrary superspins.

Let us point out that we did not need to use the constraint arising from conservation on the third point. Imposing the conservation equations at the first two points was sufficient to prove vanishing of the parity-odd contribution. Let us, however, indicate that our consideration is sensitive to the fact that the operator inserted 
at the third point is a conserved supercurrent. Indeed, we used the fact that its dimension is $s_3+1$, saturating the unitarity bound, which implies 
that this operator is a conserved supercurrent.


\subsection{Parity-even case}\label{subsection3.4}

For the parity even case, we follow the same approach. Let us denote $\mathbf{M}_{E} := \mathbf{M}_{E}^{(C)}$ (recall that the dimension of the matrix $\mathbf{M}_{E}^{(C)}$ 
in~\eqref{Parity even homogeneous system} is now $(C+1) \times (C+1)$) and consider the continuant $\D_{k}^{(C)}$, $ 1 \leq k \leq C+1$, with 
 $\D_{C+1}^{(C)} = \det[ \mathbf{M}_{E}^{(C)} ]$. The continuant satisfies eq.~\eqref{Continuant equation}, where 
%
\begin{subequations} \label{Parity even coefficients}
	\begin{align}
		a_{k} &= 2 C (3 + 2 k (4 k - 7) - 4 N (N + 2)) \nonumber\\
		& \hspace{10mm} - 
		4 (k-1) (3 + 4 k (k - 2) - 2 N (N + 2))  \, , & &1\leq k \leq C + 1 \, , \\
		b_{k} &= - 2 k (2k-1) (3 + 2N - 2k) \, , & &1\leq k \leq C \, , \\
		c_{k} &= 2 (1 + 2 C - 2 k) (1 + C - k) (1 + 2 k + 2 N) \, , & &1\leq k \leq C \, .
	\end{align}
\end{subequations}
%
Below are some examples of the matrix $\mathbf{M}_{E}$ and the determinant $\D_{C+1}^{(C)}$ for fixed $C$:
\begin{subequations} \label{Parity even matrix examples}
	\begin{align}
		\mathbf{M}_{E}^{(N,1)}= \left[ 
		\begin{array}{cc}
			-2 (2 N+1) (2 N+3) & -2 (2 N+1) \\
			2 (2 N+3) & 2 \\
		\end{array} \right]  ,
	\end{align}
	\begin{align}
		\mathbf{M}_{E}^{(N,2)} = 
		\left[
		\begin{array}{ccc}
			-4 (2 N+1) (2 N+3) & -2 (2 N+1) & 0 \\
			12 (2 N+3) & -8 \left(N^2+2 N-2\right) & -12 (2 N-1) \\
			0 & 2 (2 N+5) & 12 \\
		\end{array}
		\right] ,
	\end{align}
	\begin{align}
		\mathbf{M}_{E}^{(N,3)} = 
		\left[
		\scalemath{0.8}{\begin{array}{cccc}
			-6 (2 N+1) (2 N+3) & -2 (2 N+1) & 0 & 0 \\
			30 (2 N+3) & -2 \left(8 N^2+16 N-15\right) & -12 (2 N-1) & 0 \\
			0 & 12 (2 N+5) & -2 \left(4 N^2+8 N-39\right) & -30 (2 N-3) \\
			0 & 0 & 2 (2 N+7) & 30 \\
		\end{array}}
		\right] , 
	\end{align}
	\begin{align}
		\mathbf{M}_{E}^{(N,4)} = 
		\left[
		\scalemath{0.65}{
			\begin{array}{ccccc}
				-8 (2 N+1) (2 N+3) & -2 (2 N+1) & 0 & 0 & 0 \\
				56 (2 N+3) & -4 \left(6 N^2+12 N-11\right) & -12 (2 N-1) & 0 & 0 \\
				0 & 30 (2 N+5) & -16 \left(N^2+2 N-9\right) & -30 (2 N-3) & 0 \\
				0 & 0 & 12 (2 N+7) & -4 \left(2 N^2+4 N-45\right) & -56 (2 N-5) \\
				0 & 0 & 0 & 2 (2 N+9) & 56 \\
			\end{array}}
		\right] . 
	\end{align}
\end{subequations}
Contrary to the parity-odd case, all of the matrices above are singular, i.e. $\D_{2}^{(1)} = \D_{3}^{(2)} = \D_{4}^{(3)} = \D_{5}^{(4)} =  0$. 
For large integers we use the recursion formula \eqref{Continuant equation} to analyse the general structure of $\D_{C+1}^{(C)}$. Analogous to the parity-odd case, we computed it for arbitrary $N$ up to $C = 500$ and found in all cases $\D_{C+1}^{(C)} = 0$. 
It then follows that the matrix $\mathbf{M}_{E}^{(C)}$ is of co-rank one and 
the solution, $\vec{\a}_{E}$, to $\mathbf{M}_{E}^{(N,C)} \vec{\a}_{E} = \mathbf{0}$, is fixed up to a single overall constant in all cases, and therefore $C_{E}$ is unique.

For the parity-even case it also turns out to be possible to solve for the continuants $\D_{k}^{(C)}$ for $ 1 \leq k \leq C+1$. One can show that  $\D_{k}^{(C)}$
satisfies the following recurrence relation:
\begin{align} \label{Trick recursion - even sector}
	\D_{k}^{(C)} = \g_{k-1}^{(C)} \D_{k-1}^{(C)} \, , \hspace{5mm} \D_{1}^{(C)} = -2C (1+2 N) (3+2 N) \, , \hspace{5mm} 1 \leq k \leq C + 1 \, ,
\end{align}
where $\g^{(C)}_{k}$ is given by 
%
\begin{align} \label{e1}
	\g^{(C)}_{k} = - 2 (C-k) (1 + 2 N - 2k) (3 + 2N + 2k ) \, .
\end{align}
%
To show this, consider again the combination $\D_{k+2}^{(C)} - a_{k+2} \D_{k+1}^{(C)} + b_{k+1} c_{k+1} \D_{k}^{(C)}$. 
Using eqs. \eqref{Trick recursion - even sector}, \eqref{e1}, we obtain $( \g_{k+1} \g_{k} - a_{k+2} \g_{k}  + b_{k+1} c_{k+1} ) \D^{(C)}_{k}$. 
It is then simple to show using eqs.~\eqref{Parity even coefficients} and~\eqref{e1} that this combination vanishes for arbitrary $k, N, C$, which implies that the recurrence relation~\eqref{Continuant equation} is indeed satisfied. 
Analogous to the parity-odd case, it is possible to find an explicit solution for the recurrence relation~\eqref{Trick recursion - even sector}, and we find the following general formula for the continuant:
\begin{align} \label{Even continuant}
	\D_{k}^{(C)} =  (-1)^{N+1} \frac{ 2^{3k} C! }{ \pi \, (C-k)! } \, \Gamma(\tfrac{1}{2} - N + k ) \, \Gamma( \tfrac{3}{2} + N + k )  \, , \hspace{5mm} 1 \leq k \leq C \,.
\end{align}
%
For $k = C+1$, we have $\Delta_{C+1}^{(C)} = \g_{C}^{(C)} \Delta_{C}^{(C)}$. However, we note that $\g_{C}^{(C)} = 0$, which implies 
$\Delta_{C+1}^{(C)} = \det[ \mathbf{M}_{E}^{(C)} ] = 0$, for arbitrary $N, C$. Hence, we have shown that the matrix $\mathbf{M}_{E}$ is always of co-rank one and
the system~\eqref{Parity even homogeneous system} has a unique non-trivial solution.\footnote{Alternatively, note that \eqref{Even continuant} implies that the largest non-trivial minor of $\mathbf{M}_{E}$ is of dimension $C \times C$. Therefore, $\text{Rank}(\mathbf{M}_{E}) = C$, which implies $\text{Nullity}( \mathbf{M}_{E} ) = 1$.} This, in turn, implies that the tensor $C_{E}(X; u,v)$ is 
unique up to an overall coefficient. The explicit solution to the system $\mathbf{M} \vec{\a} = \mathbf{0}$ and, thus, for the tensor $C_E$ 
can also be found for arbitrary $N,C$. Indeed, by analysing the nullspace of $\mathbf{M}_{E}$ we obtain the following solution for the coefficients $\a_{2k}$ of \eqref{zh7}:
\begin{align}
	\a_{2k} = \frac{ (-1)^k 2^{2 k} \, \Gamma(C+1) \, \Gamma(k+N+\tfrac{3}{2}) }{ \Gamma(2k+1) \, \Gamma(C-k+1) \, \Gamma(N+\tfrac{3}{2}) } \, \a_{0} \, , \hspace{10mm} 1 \leq k \leq C \, .
\end{align}
This is also a solution to the parity-even sector of the recurrence relations \eqref{Linear homogeneous equations}, which can be explicitly checked. Hence, we have obtained a unique solution (up to an overall coefficient) for $C_{E}$ in explicit form for arbitrary superspins.


Now recall that the three-point functions under consideration are determined not just by the tensor $C^{(r_{1} ... r_{N+1})(k_{1} ... k_{C})}$  
but also by the tensor $D_{S}^{(r_{1} ... r_{N+1})(k_{1} ... k_{C+1})}$ which is transverse (see eq.~\eqref{e4}) and related to $C^{(r_{1} ... r_{N+1})(k_{1} ... k_{C})}$  
by the algebraic relation~\eqref{C and D algebraic relations}. Remarkably, it is possible to solve for $D_S$ in terms of $C$ using~\eqref{C and D algebraic relations}. 
To show it let us begin by constructing irreducible decompositions for $C$ and $D$. Since we know that $C$ is parity-even, it cannot contain $\e$, and we use the decompositions
\begin{align} \label{C irreducible decomposition}
	C^{(r_{1} ... r_{N+1})(k_{1} ... k_{C})} = C_{1}^{(r_{1} ... r_{N+1} k_{1} ... k_{C})} + \sum_{i=1}^{N+1} \sum_{j=1}^{C} \eta^{r_{i} k_{j}} C_{2}^{(r_{1} ... \hat{r}_{i} ... r_{N+1} k_{1} ... \hat{k}_{j} ... k_{C})} \nonumber \\ 
	+ \sum_{j > i=1}^{N+1} \eta^{r_{i} r_{j}} C_{3}^{(r_{1} ... \hat{r}_{i} \hat{r}_{j} ... r_{N+1} k_{1} ... k_{C})}
	+ \sum_{j > i=1}^{C} \eta^{k_{i} k_{j}} C_{4}^{(r_{1} ... r_{N+1} k_{1} ... \hat{k}_{i} \hat{k}_{j} ... k_{C})} \, ,
\end{align}
where $C_{2}, C_{3}, C_{4}$ are the irreducible components of rank $N+C-1$ ($C_{4}$ exists only for $C > 1$). 
Requiring that the above ansatz is traceless in the appropriate groups of indices fixes $C_{3}$ and $C_{4}$ in terms of $C_{2}$ as follows (indices suppressed):
\begin{align}
	C_{3} = - \frac{2C}{2N+1} \, C_{2} \, , \hspace{15mm} C_{4} = - \frac{2(N+1)}{2C-1} \, C_{2} \, .
\end{align}
Hence, $C$ is determined completely in terms of the totally symmetric and traceless tensors $C_{1}$ and $C_{2}$. Now let us construct an irreducible decomposition of $D_{S}$. 
Due to the algebraic relation \eqref{C and D algebraic relations - more symmetric}, we know that $D_{S}$ must be linear in $\e$. 
The only way to construct $D_{S}$ such that it contains $\e$ is by using the following decomposition:
\begin{equation} \label{D_{S} irreducible decomposition}
	D_{S}^{(r_{1} ... r_{N+1})(k_{1} ... k_{C+1})} = \sum_{i=1}^{N+1} \sum_{j=1}^{C+1} \e^{q r_{i} k_{j}} T^{q, (r_{1} ... \hat{r}_{i} ... r_{N+1} k_{1} ... \hat{k}_{j} ... k_{C+1})} \, ,
\end{equation}
where the tensor $T$ (of rank $N+C+1$) is decomposed as follows:
\begin{align}
	T^{q, (r_{1} ... r_{N})(k_{1} ... k_{C})} = T_{1}^{(r_{1} ... r_{N} k_{1} ... k_{C} q)} + \sum_{i=1}^{N} \sum_{j=1}^{C} \eta^{r_{i} k_{j}} T_{2}^{(r_{1} ... \hat{r}_{i} ... r_{N} k_{1} ... \hat{k}_{j} ... k_{C} q)} \nonumber \\ 
	+ \sum_{j > i=1}^{N} \eta^{r_{i} r_{j}} T_{3}^{(r_{1} ... \hat{r}_{i} \hat{r}_{j} ... r_{N} k_{1} ... k_{C} q)}
	+ \sum_{j > i=1}^{C} \eta^{k_{i} k_{j}} T_{4}^{(r_{1} ... r_{N} k_{1} ... \hat{k}_{i} \hat{k}_{j} ... k_{C} q)} \, .
	\label{zh3}
\end{align}
Here $T_1$ is the irreducible component of rank $N+C+1$ and $T_{2}, T_{3}, T_{4}$ are the irreducible components of rank $N+C-1$ (where $T_{4}$ exists only for $C > 1$). It should be noted that one could also consider contributions to $T$ proportionate to $\eta^{q r_{i}}$, $\eta^{q k_{j}}$, 
but such contributions will cancel when substituted into \eqref{D_{S} irreducible decomposition} and, hence, they do not contribute to the irreducible decomposition of $D_S$.  
Requiring that $D$ is traceless on the appropriate groups of indices fixes $T_{3}$ and $T_{4}$ in terms of $T_{2}$ as follows (indices suppressed):
\begin{align}
	T_{3} = - \frac{2C}{2N+1} \, T_{2} \, , \hspace{15mm} T_{4} = - \frac{2N}{2C-1} \, T_{2} \, .
	\label{zh4}
\end{align}
Hence, $D_{S}$ is described completely in terms of the totally symmetric and traceless tensors $T_{1}$ and $T_{2}$. If we now consider the algebraic relations \eqref{C and D algebraic relations - more symmetric} and substitute in the above decompositions, after some tedious calculation one obtains:
\begin{align}
	0 &= \eta_{r_{1} k_{1}} D_{S}^{(r_{1} ... r_{N+1})(k_{1} ... k_{C+1})} + \frac{1}{C+1} \sum_{i=2}^{C+1} \e^{k_{i}}{}_{r_{1} k_{1}} C^{ (r_{1} ... r_{N+1})( k_{1} k_{2} ... \hat{k}_{i} ... k_{C+1} ) } \nonumber \\
	&= \sum_{i=2}^{N+1} \sum_{j=2}^{C+1} \e^{q r_{i} k_{j}} \Big\{ \t(N,C) \, T_{2}^{(\hat{r}_{1} ... \hat{r}_{i} ... r_{N+1} \hat{k}_{1} ... \hat{k}_{j} ... k_{C+1} q)}  \\
	& \hspace{30mm} + \frac{1}{C+1} \bigg( 1 + \frac{2C}{2N + 1} \bigg) C_{2}^{(\hat{r}_{1} ... \hat{r}_{i} ... r_{N+1} \hat{k}_{1} ... \hat{k}_{j} ... k_{C+1} q)} \Big\}\,, \nonumber
\end{align}
where the constant $\t(N,C)$ is defined as follows:
\begin{equation}
	\t(N,C) = \frac{ 6 C^2 + 2 N^2 + 3 C (1 + 4 N) - 5 N - 3 }{(2 C - 1) (1 + 2 N)} \, .
\end{equation}
Requiring that the above combination vanishes gives $T_{2} = \xi_{2} \, C_{2}$, where
\begin{equation}
	\xi_{2} = - \frac{ (2 C - 1) (1 + 2 C + 2 N) }{ (1 + C) ( 6 C^2 + 2 N^2 + 3 C (1 + 4 N) - 5 N - 3 ) } \, .
\end{equation}
One can proceed in a similar way with the second algebraic relation in \eqref{C and D algebraic relations - more symmetric}. After some calculation we found
that it relates $T_1$ and $C_1$ as  $T_{1} = \xi_{1} \, C_{1}$, where
\begin{equation}
	\xi_{1} = - \frac{ 2 C + 1 }{ (C+1)(N+C+2)} \, .
\end{equation}
A similar calculation was performed in \cite{Buchbinder:2015qsa}, where it was shown that for $A = B = C = 1$, $\xi_{1} = - \frac{3}{10}$, $\xi_{2} = - \frac{1}{8}$, 
which is in full agreement with the expressions above. 

Finally, we need to show that $D_S$ found this way is transverse, i.e.
\begin{equation}
\pa_{p} D_{S}^{(p r_1 \dots r_N) (k_1 \dots k_{C+1})} = 0\,. 
\label{e5}
\end{equation}
For this let us define the tensor 
\begin{equation}
E^{( r_1 \dots r_N) (k_1 \dots k_{C+1})}= \pa_{p} D_{S}^{(p r_1 \dots r_N) (k_1 \dots k_{C+1})} \, . 
\end{equation}
Our aim is to show that $E^{( r_1 \dots r_N) (k_1 \dots k_{C+1})}=0$. To find $E^{( r_1 \dots r_N) (k_1 \dots k_{C+1})}$ we contract $D_S$ in 
eqs.~\eqref{D_{S} irreducible decomposition}, \eqref{zh3}, \eqref{zh4}  with the derivative. 
However, from the algebraic relations~\eqref{C and D algebraic relations} and the fact that $C$ is transverse it follows that $E^{( r_1 \dots r_N) (k_1 \dots k_{C+1})}$ is totally symmetric and traceless:
\begin{equation}
	E^{(r_{1} ... r_{N})(k_{1} ... k_{C+1})} \equiv E^{(r_{1} ... r_{N} k_{1} ... k_{C+1})}  \, .
\end{equation}
Using eqs.~\eqref{D_{S} irreducible decomposition}, \eqref{zh3}, \eqref{zh4} we find that the totally symmetric and traceless contribution is given by 
\begin{align}
E^{(r_{2} ... r_{N+1} k_{1} ... k_{C+1})} &= -\sum_{i=2}^{N+1} \e_{ p q }{}^{ r_{i} } \pa^{p} T_{1}^{(q r_{2} ... \hat{r}_{i} ... r_{N+1} k_{1} ... k_{C+1} )} \nonumber  \\
& \hspace{15mm} -\sum_{j=1}^{C+1} \e_{ p q }{}^{ k_{j} } \pa^{p} T_{1}^{(q r_{2} ... r_{N+1} k_{1} ... \hat{k}_{j} ... k_{C+1})} \, .
\label{zh6}
\end{align}
Since the tensor $T_1$ is proportional to $C_1$, it is constructed out of the vector $X^m$ and the Minkowski metric $\eta^{mn}$. It is not difficult to see that for
a tensor $T_{1}$ constructed out of $X^m$ and $\eta^{mn}$, the combination \eqref{zh6} must vanish. Hence $E = 0$ and $D_{S}$ is transverse.

Let us summarise the results of this subsection. We have shown that the parity-even contribution is fixed up to an overall coefficient. Moreover, it can be explicitly found 
for arbitrary superspins by solving a linear homogeneous system of equations with the tri-diagonal matrix \eqref{Parity even homogeneous system}. 
Once the solution for the parity-even coefficients $\alpha_{2k}$ is found, the tensor $C$ is obtained using eq.~\eqref{zh7} and the tensor $D_S$ is obtained 
from $C$ as discussed above. Note that our analysis for the parity even solution is incomplete since we have not imposed the conservation condition at the third point. 
This is technically difficult to impose using the approach outlined in the present paper, and from this viewpoint the computational approach developed 
in~\cite{Buchbinder:2023fqv} is far more useful. However, it was verified in~\cite{Buchbinder:2023fqv} up to $s_{i} = 20$ that if $s_3 \leq s_1+ s_2$ 
(that is, the third triangle inequality is satisfied) then the third conservation equation is automatically satisfied and does not result in any new restrictions on the three-point function. 
The analysis in this subsection assumes that this property continues to hold for arbitrary superspins.

\subsection{Point-switch symmetries}

For the $\langle \mathbf{J}^{}_{F} \mathbf{J}'_{F} \mathbf{J}''_{F} \rangle$ three-point functions we can also examine the case where $\mathbf{J} = \mathbf{J}'$ for arbitrary superspins. We want to determine the conditions under which the parity-even solution satisfies the point-switch symmetry. If we consider the condition \eqref{Point-switch 1} and the irreducible decomposition \eqref{FFF - irreducible decomposition}, we obtain the following conditions on $C$ and $D$:
\begin{subequations}
	\begin{align}
		C_{E}^{(r_{1} ... r_{N+1})(k_{1} ... k_{C})}(X) - C_{E}^{(r_{1} ... r_{N+1})(k_{1} ... k_{C})}(-X) &= 0 \, , \label{Point switch 1 - C} \\
		D_{E}^{(r_{1} ... r_{N+1})(k_{1} ... k_{C+1})}(X) - D_{E}^{(r_{1} ... r_{N+1})(k_{1} ... k_{C+1})}(-X) &= 0 \, . \label{Point switch 1 - D}
	\end{align}
\end{subequations}
Let us consider \eqref{Point switch 1 - C} first. Using auxiliary spinors, this condition may be written as
\begin{align} \label{Point switch 1 - C in auxiliary spinors}
	C_{E}( X; u,v ) - C_{E}(- X; u,v ) = 0 \, .
\end{align}
However, recall that $C_{E}(X; u,v)$ is of the form
\begin{align}
	C_{E}(X; u,v ) &= X^{C-N-2} \sum_{k=0}^{C} \a_{2k} P_{3}^{2k} Q_{3}^{2(C - k)} Z_{1}^{N-C+1} \, .
\end{align}
For $\mathbf{J} = \mathbf{J}'$, we have $N = A + B = 2A$, and from \eqref{Point switch 1 - C in auxiliary spinors} we obtain
\begin{equation}
	\sum_{k=0}^{C} (1 + (-1)^{C}) \, \a_{2k} P_{3}^{2k} Q_{3}^{2(C - k)} Z_{1}^{N-C+1} = 0 \, .
\end{equation}
Hence, the parity-even solution $C_{E}(X; u, v)$ satisfies the point-switch only for $C$ odd, i.e. for $s_{3} = 2k + \tfrac{3}{2}$, $k \in \mathbb{Z}_{\geq 0}$, 
which is consistent with the results of \cite{Buchbinder:2023fqv}. 

\newpage

Now assume that $C_{E}$ satisfies the point-switch symmetry \eqref{Point switch 1 - C}. 
We want to show that $D_{E}$, which is fully determined by $C_{E}$, satisfies \eqref{Point switch 1 - D}. Since \eqref{Point switch 1 - C} is satisfied, 
from \eqref{C irreducible decomposition} we must have $C_{1}(X) = C_{1}(-X)$, $C_{2}(X) = C_{2}(-X)$ (indices suppressed). 
Now consider \eqref{D decomposition - more symmetric} and the irreducible decomposition for $D_{S}$ given by \eqref{D_{S} irreducible decomposition}. 
Since $T_{1} \propto C_{1}$, $T_{2} \propto C_{2}$, we have $T_{1}(X) = T_{1}(-X)$, $T_{2}(X) = T_{2}(-X)$. It is then easy to see by 
substituting \eqref{D decomposition - more symmetric}, \eqref{D_{S} irreducible decomposition} into \eqref{Point switch 1 - D} that $D_{E}$ also 
satisfies the point-switch symmetry. Concerning point-switch symmetries that involve $\mathbf{J}''$, these are difficult to check using the approach 
outlined in this paper, as one must compute $\tilde{\cH}$ using \eqref{H tilde} and check \eqref{Point-switch 2}. However, the results of \cite{Buchbinder:2023fqv} cover 
these cases in more detail and so we will not discuss them here. 


\section{Grassmann-odd three-point functions -- \texorpdfstring{$\langle \mathbf{J}^{}_{F} \mathbf{J}'_{B} \mathbf{J}''_{B} \rangle$
	}{< F B B >}}\label{section4}

Let us now consider the case $\langle \mathbf{J}^{}_{F} \mathbf{J}'_{B} \mathbf{J}''_{B} \rangle$, which proves to be considerably simpler. 
Let us begin with making important comments on the arrangement of the operators in this three-point function. First, we arrange them in such a way that 
the operator at the third position is bosonic. Second, we arrange them in such a way that the third superspin satisfies the triangle inequality, that is  $s_3 \leq s_1+ s_2$. 
As was mentioned in the previous section if one of the triangle inequalities is violated the remaining two are necessarily satisfied. This means that we can always 
place a bosonic operator with the superspin satisfying the triangle inequality at the third position. 

We consider one Grassmann-odd current, $\mathbf{J}_{\a(2A+1)}$, of spin $s_{1} = A+ \tfrac{1}{2}$, and two Grassmann-even currents $\mathbf{J}'_{\a(2B)}$, $\mathbf{J}''_{\g(2C)}$, of spins $s_{2} = B$, $s_{3} = C$ respectively, where $A,B,C$ are positive integers. All information about the correlation function
\begin{equation}
	\langle \, \mathbf{J}^{}_{\a(2A+1)}(z_{1}) \, \mathbf{J}'_{\b(2B)}(z_{2}) \, \mathbf{J}''_{\g(2C)}(z_{3}) \rangle \, ,
\end{equation}
is now encoded in a homogeneous tensor field $\cH_{\a(2A+1) \b(2B+1) \g(2C+1)}(\boldsymbol{X}, \Q)$, which satisfies the scaling property
\begin{equation}
	\cH_{\a(2A+1) \b(2B) \g(2C)}( \l^{2} \boldsymbol{X}, \l \Q) = (\l^{2})^{C-A-B-\tfrac{3}{2}}\cH_{\a(2A+1) \b(2B) \g(2C)}(\boldsymbol{X}, \Q) \, .
\end{equation}
Analogous to the previous case, for each set of totally symmetric spinor indices (the $\a$'s, $\b$'s and $\g$'s respectively), we convert pairs of them into vector indices as follows:
\begin{align}
	\cH_{\a(2A+1) \b(2B) \g(2C)}(\boldsymbol{X}, \Q) &\equiv \cH_{\a \a(2A), \b(2B), \g(2C)}(\boldsymbol{X}, \Q) \nonumber \\
	&= (\g^{m_{1}})_{\a_{1} \a_{2}} ... (\g^{m_{A}})_{\a_{2A-1} \a_{2A}} \nonumber \\
	& \times (\g^{n_{1}})_{\b_{1} \b_{2}} ... (\g^{n_{B}})_{\b_{2B-1} \b_{2B}} \nonumber \\
	& \times (\g^{k_{1}})_{\g_{1} \g_{2}} ... (\g^{k_{C}})_{\g_{2C-1} \g_{2C}} \nonumber \\
	& \times \cH_{\a, m_{1} ... m_{A}, n_{1} ... n_{B}, k_{1} ... k_{C} }(\boldsymbol{X}, \Q) \, .
\end{align}
Again, the equality above holds only if and only if $\cH_{\a, m_{1} ... m_{A} n_{1} ... n_{B} k_{1} ... k_{C} }(\boldsymbol{X}, \Q)$ is totally symmetric and traceless in each group of vector indices. It is also required that $\cH$ is subject to the $\g$-trace constraint
\begin{align}
	(\g^{m_{1}})_{\s}{}^{\a} \cH_{\a, m_{1} ... m_{A}, n_{1} ... n_{B}, k_{1} ... k_{C} }(\boldsymbol{X}, \Q) = 0 \, . \label{FBB - Gamma trace 1}
\end{align}
Indeed, since $\cH$ is Grassmann-odd it is linear in $\Q$, and we decompose $\cH$ it follows (again, raising all indices for convenience):
\begin{subequations} \label{FBB - irreducible decomposition}
\begin{equation}
	\cH^{\a, m(A) n(B) k(C) }(\boldsymbol{X}, \Q) = \sum_{i=1}^{2} \cH_{i}^{\a, m(A) n(B) k(C) }(X, \Q) \, , 
\end{equation}
\vspace{-5mm}
\begin{align}
	\cH_{1}^{\a, m(A) n(B) k(C) }(X, \Q) &= \Q^{\a} A^{m(A) n(B) k(C)}(X) \, , \\
	\cH_{2}^{\a, m(A) n(B) k(C) }(X, \Q) &= (\gamma_{p})^{\a}{}_{\d} \Q^{\d} B^{p, m(A) n(B) k(C)}(X) \, .
\end{align}
\end{subequations}
Hence, in this case there are only two contributions to consider. The conservation equations \eqref{Conservation equations on H - 1}, \eqref{Conservation equations on H - 2} are now equivalent to the following constraints on $\cH$ with vector indices:
\begin{subequations}
	\begin{align}
		\cD_{\a} \cH^{\a, m(A) n(B) k(C) }(\boldsymbol{X}, \Q) = 0 \, , \\
		(\gamma_{n})_{\s}{}^{\b} \cQ_{\b} \cH^{\a, m(A) n n(B-1) k(C) }(\boldsymbol{X}, \Q) = 0 \, .
	\end{align}
\end{subequations}
They split up into constraints $O(\Q^{0})$
\begin{subequations}
	\begin{align}
		\frac{\pa}{\pa \Q^{\a}} \cH^{\a, m(A) n(B) k(C) }(\boldsymbol{X}, \Q) &= 0 \, , \label{FBB - conservation equations O(0) - 1} \\
		(\gamma_{n})_{\s}{}^{\b} \frac{\pa}{\pa \Q^{\b}} \cH^{\a, m(A) n n(B-1) k(C) }(\boldsymbol{X}, \Q) &= 0 \, , \label{FBB - conservation equations O(0) - 2}
	\end{align}
\end{subequations}
and $O(\Q^{2})$
\begin{subequations}
	\begin{align}
		(\g^{m})_{\a \d} \Q^{\d} \frac{\pa}{\pa X^{m}} \cH^{\a, m(A) n(B) k(C) }(\boldsymbol{X}, \Q) &= 0 \, , \label{FBB - conservation equations O(2) - 1} \\
		(\gamma_{n})_{\s}{}^{\b} (\g^{m})_{\b \d} \Q^{\d} \frac{\pa}{\pa X^{m}} \cH^{\a, m(A) n n(B-1) k(C) }(\boldsymbol{X}, \Q) &= 0 \, . \label{FBB - conservation equations O(2) - 2}
	\end{align}
\end{subequations}
Using the irreducible decomposition \eqref{FBB - irreducible decomposition}, equation \eqref{FBB - conservation equations O(0) - 1} immediately results in $A = 0$, while \eqref{FBB - conservation equations O(0) - 2} gives
\begin{align} \label{FBB - B constraints 1}
	\eta_{pn} B^{p, m(A) n n(B-1) k(C) } &= 0 \, , & \e_{qpn} B^{p,m(A) n n(B-1) k(C)} &= 0 \, .
\end{align}
Next, after imposing the $\g$-trace condition \eqref{FBB - Gamma trace 1}, we find that $B$ must satisfy
\begin{align} \label{FBB - B constraints 2}
	\eta_{pm} B^{p, m m(A-1) n(B) k(C) } &= 0 \, , & \e_{qpm} B^{p, m m(A-1) n(B) k(C)} &= 0 \, .
\end{align}
Altogether \eqref{FBB - B constraints 1} and \eqref{FBB - B constraints 2} imply that $B$ is symmetric and traceless in the indices $p$, $m_{1}, ..., m_{A}$, $n_{1}, ..., n_{B}$, i.e.
\begin{equation}
	B^{p, m(A) n(B) k(C)} \equiv B^{(p m(A) n(B)), k(C)} \, .
\end{equation}
If we now consider the equations arising from conservation at $O(\Q^{2})$, a simple computation shows that $B$ must satisfy
\begin{align}
	\pa_{p} B^{(p m(A) n(B)), k(C)} = 0 \, .
\end{align}
Therefore we need to construct a single transverse tensor $B$ of rank $A+B+C+1$. We see that the tensor $B$ has absolutely same properties as the tensor $C$ from the previous 
section. Hence, the analysis becomes exactly the same as for  $C$ in the $\langle \mathbf{J}^{}_{F} \mathbf{J}'_{F} \mathbf{J}''_{F} \rangle$ case and we will not repeat it. 
Recalling the results from the previous section, we find that $\langle \mathbf{J}^{}_{F} \mathbf{J}'_{B} \mathbf{J}''_{B} \rangle$ has vanishing parity-odd contribution 
for all values of the superspins and a unique parity-even structure.\footnote{Here we also have assumed that with our arrangement of the operators 
the conservation condition at the third point is automatically satisfied for all superspins. It is verified in our computational approach in~\cite{Buchbinder:2023fqv} up to $s_i=20$.}
The explicit form of the parity-even solution can be found from the tri-diagonal system of linear equations just like in the previous section.


\section*{Acknowledgements}
The authors are grateful to Sergei Kuzenko for valuable discussions. The work of E.I.B. is supported in part by the Australian Research Council, projects DP200101944
and DP230101629. The work of B.S. is supported by the \textit{Bruce and Betty Green Postgraduate Research Scholarship} under the Australian Government Research Training Program.


\vfill


\appendix

\section{3D conventions and notation}\label{AppA}

For the Minkowski metric we use the ``mostly plus'' convention: $\eta_{mn} = \text{diag}(-1,1,1)$. Spinor indices are then raised and lowered with the $\text{SL}(2,\mathbb{R})$ invariant anti-symmetric $\varepsilon$-tensor
\begin{subequations}
	\begin{align}
		\ve_{\a \b} = 
		\begingroup
		\setlength\arraycolsep{4pt}
		\begin{pmatrix}
			\, 0 & -1 \, \\
			\, 1 & 0 \,
		\end{pmatrix}
		\endgroup 
		\, , & \hspace{6mm}
		\ve^{\a \b} =
		\begingroup
		\setlength\arraycolsep{4pt}
		\begin{pmatrix}
			\, 0 & 1 \, \\
			\, -1 & 0 \,
		\end{pmatrix}
		\endgroup 
		\, , \hspace{6mm}
		\ve_{\a \g} \ve^{\g \b} = \d_{\a}{}^{\b} \, , \\[4mm]
		& \hspace{-8mm} \f_{\a} = \ve_{\a \b} \, \f^{\b} \, , \hspace{12mm} \f^{\a} = \ve^{\a \b} \, \f_{\b} \, .
	\end{align}
\end{subequations}
The $\g$-matrices are chosen to be real, and are expressed in terms of the Pauli matrices, $\s$, as follows:
\begin{subequations}
	\begin{align}
		(\g_{0})_{\a}{}^{\b} = - \text{i} \s_{2} = 
		\begingroup
		\setlength\arraycolsep{4pt}
		\begin{pmatrix}
			\, 0 & -1 \, \\
			\, 1 & 0 \,
		\end{pmatrix}
		\endgroup 
		\, , & \hspace{8mm}
		(\g_{1})_{\a}{}^{\b} = \s_{3} = 
		\begingroup
		\setlength\arraycolsep{4pt}
		\begin{pmatrix}
			\, 1 & 0 \, \\
			\, 0 & -1 \,
		\end{pmatrix}
		\endgroup 
		\, , \\[3mm]
		(\g_{2})_{\a}{}^{\b} = - \s_{1} &= 
		\begingroup
		\setlength\arraycolsep{4pt}
		\begin{pmatrix}
			\, 0 & -1 \, \\
			\, -1 & 0 \,
		\end{pmatrix}
		\endgroup 
		\, ,
	\end{align}
	\vspace{1mm}
	\begin{equation}
		(\g_{m})_{\a \b} = \ve_{\b \d} (\g_{m})_{\a}{}^{\d} \, , \hspace{10mm} (\g_{m})^{\a \b} = \ve^{\a \d} (\g_{m})_{\d}{}^{\b} \, .
	\end{equation}
\end{subequations}
The $\g$-matrices are traceless and symmetric
\begin{equation}
	(\g_{m})^{\a}{}_{\a} = 0 \, , \hspace{10mm} (\g_{m})_{\a \b} = (\g_{m})_{\b \a} \, ,
\end{equation} 
and also satisfy the Clifford algebra
\begin{equation}
	\g_{m} \g_{n} + \g_{n} \g_{m} = 2 \eta_{mn} \, .
\end{equation}
For products of $\g$-matrices we make use of the identities
\begin{subequations}
	\begin{align}
		(\g_{m})_{\a}{}^{\r} (\g_{n})_{\r}{}^{\b} &= \eta_{mn} \d_{\a}{}^{\b} + \e_{mnp} (\g^{p})_{\a}{}^{\b} \, , \\[2mm]
		(\g_{m})_{\a}{}^{\r} (\g_{n})_{\r}{}^{\s} (\g_{p})_{\s}{}^{\b} &= \eta_{mn} (\g_{p})_{\a}{}^{\b} - \eta_{mp} (\g_{n})_{\a}{}^{\b} + \eta_{np} (\g_{m})_{\a}{}^{\b} + \e_{mnp} \d_{\a}{}^{\b} \, ,
	\end{align}
\end{subequations}
where we have introduced the 3D Levi-Civita tensor $\e$, with $\e^{012} = - \e_{012} = 1$. We also have the orthogonality and completeness relations for the $\g$-matrices
\begin{equation}
	(\g^{m})_{\a \b} (\g_{m})^{\r \s} = - \d_{\a}{}^{\r} \d_{\b}{}^{\s}  - \d_{\a}{}^{\s}  \d_{\b}{}^{\r} \, , \hspace{8mm} (\g_{m})_{\a \b} (\g_{n})^{\a \b} = -2 \eta_{mn} \, .
\end{equation}
The $\g$-matrices are used to swap from vector indices to spinor indices. For example, given some three-vector $x_{m}$, it may equivalently be expressed in terms of a symmetric second-rank spinor $x_{\a \b}$ as follows:
\begin{subequations}
	\begin{align}
		x_{\a \b} = (\g^{m})_{\a \b} x_{m}  \, , \hspace{5mm} x_{m} = - \frac{1}{2} (\g_{m})^{\a \b} x_{\a \b} \, , \\[2mm]
		\det (x_{\a \b}) = \frac{1}{2} x^{\a \b} x_{\a \b} = - x^{m} x_{m} = -x^{2} \, .
	\end{align}
\end{subequations}
The same conventions are also adopted for the spacetime partial derivatives $\partial_{m}$
\begin{subequations}
	\begin{align}
		\partial_{\a \b} = (\g^{m})_{\a \b} \partial_{m}  \, , \hspace{5mm} \partial_{m} = - \frac{1}{2} (\g_{m})^{\a \b} \partial_{\a \b} \, , \\[2mm]
		\partial_{m} x^{n} = \d_{m}^{n} \, , \hspace{5mm} \partial_{\a \b} x^{\r \s} = - \d_{\a}{}^{\r} \d_{\b}{}^{\s}  - \d_{\a}{}^{\s}  \d_{\b}{}^{\r} \, ,
	\end{align}
\end{subequations}
\begin{equation}
	\x^{m} \partial_{m} = - \frac{1}{2} \x^{\a \b} \partial_{\a \b} \, .
\end{equation}
We also define the supersymmetry generators $Q_{\a}$
\begin{equation}
	Q_{\a} = \text{i} \frac{\partial}{\partial \q^{\a}} + (\g^{m})_{\a \b} \q^{\b} \frac{\partial}{\partial x^{m}} \, , \label{Supercharges}
\end{equation}
and the covariant spinor derivatives
\begin{equation}
	D_{\a} = \frac{\partial}{\partial \q^{\a}} + \text{i} (\g^{m})_{\a \b} \q^{\b} \frac{\partial}{\partial x^{m}} \, , \label{Covariant spinor derivatives}
\end{equation}
which anti-commute with the supersymmetry generators, $\{ Q_{\a} , D_{\b}\} = 0$, and obey the standard anti-commutation relations
\begin{equation}
	\{ D_{\a} , D_{\b} \} = 2 \text{i} \, (\g^{m})_{\a \b} \partial_{m} \, .
\end{equation}



\printbibliography[heading=bibintoc,title={References}]



\end{document}